\documentclass[aps,prd,nofootinbib,amsmath,amssymb]{revtex4}
\usepackage{eurosym}
\usepackage{amsmath,amssymb}
\usepackage{graphicx}
\usepackage{float}
\usepackage{dcolumn}
\usepackage{bm}
\usepackage{color}
\usepackage{multirow}

\usepackage{array}
\usepackage{booktabs}
\usepackage[section]{placeins}
\usepackage[detect-all]{siunitx}
\usepackage{makecell}
\usepackage{xcolor}
\usepackage{yfonts}

\graphicspath{{./}}

\newcolumntype{P}[1]{>{\centering\arraybackslash}p{#1}}

\begin{document}

\title{Lightest $\Omega$-baryon systems with strangeness $-3$ and $-6$ within the framework of Faddeev and Faddeev--Yakubovsky equations}

\author{R. Lazauskas$^{1}$, R. Ya. Kezerashvili$^{2,3}$, I. Filikhin$^{4}$}
\affiliation{$^{1}$IPHC, CNRS/IN2P3, Universit´e de Strasbourg, 67037 Strasbourg, France\\
$^{2}$New York City College of Technology, The City University of New York,
Brooklyn, NY, USA\\
$^{3}$The Graduate School and University Center, The City University of New
York, New York, NY, USA\\
$^{4}$North Carolina Central University, Durham, NC, USA 
}

\begin{abstract}
We investigate the binding energies and spatial properties
of few-body $\Omega$-baryon systems with strangeness
$S=-3$ and $S=-6$ using the configuration-space Faddeev
and Faddeev--Yakubovsky equations.
For the first time, microscopic Faddeev--Yakubovsky
calculations are performed for the four- and five-body
$\Omega NNN$, $\Omega NNNN$, and $\Omega\Omega NN$
systems.
We employ HAL QCD and meson-exchange $\Omega N$
interactions, together with the HAL QCD $\Omega\Omega$
potential, and examine the effects of Coulomb interactions
and spin--isospin configurations.
The calculations predict substantial enhancement of binding
with increasing baryon number, with the most strongly bound
$\Omega NNNN$ configurations reaching approximately
$60$~MeV and bound $\Omega\Omega NN$ states in the
$S=-6$ sector.
The results reveal pronounced spin dependence and compact
few-body structures, with the $\Omega$ baryon preferentially
localized near the center-of-mass.
These findings provide new predictions for light
multistrange nuclear systems and extend microscopic
few-body calculations into the four- and five-body
$\Omega$-baryon sectors.
\end{abstract}

\date{\today }

\maketitle

\date{\today }

\section{Introduction}

The $\Omega$-baryons form a family of hadrons that can carry electric charges of  $+2$, $+1$, or $-1$. The negatively charged state, $\Omega^{-}(sss)$,  is composed of three strange quarks \cite{B64,DataGroup,2009}, carries strangeness $S=-3$, and has a rest mass of 1672.45 MeV/$c^{2}$ \cite{Partlist2012}. Systems containing strange quarks are of considerable interest in nuclear physics, as they provide a means of exploring the strong interaction beyond the sector of ordinary nucleons and $\phi$-mesic nuclear system with hidden strangeness \cite{Metag2017}. In particular, the interaction of the $\Omega$ baryon with nucleons plays an important role in understanding the strong interaction in systems involving heavy strange quarks.

The $\Omega^{-}$ baryon is of particular interest because it contains no light ($u,d$) quarks and exhibits a comparatively long lifetime relative to other hyperons. Its quark content additionally reduces Pauli-blocking effects at quark-level, rendering the $\Omega$N interaction a sensitive probe for the dynamics of strongly interacting systems. Accordingly, this interaction has been investigated extensively, including in the context of possible deeply bound $\Omega N$ states. Such scenarios are further motivated by advances in lattice QCD calculations, which suggest a pronounced attractive component in specific spin channels.

The possibility of a bound $\Omega N$ dibaryon has been explored in a variety of quark-model calculations \cite{Goldman1987,Oka1988,Li2000,Pang2004}. An $\Omega N$ bound state with strangeness $S=-3$ was first proposed in Ref.~\cite{Goldman1987}, based on the potential quark model and the MIT bag model. Subsequent investigations considered candidate S-wave dibaryons with different strangeness quantum numbers, including the $\Omega N$ system, within the chiral SU(3) quark model \cite{Li2000}. In particular, the $S=-3$  six-quark system was studied using the resonating-group method in the single $\Omega N$ channel with a spin $S=2$. The effective $\Omega N$ interaction was also investigated within the refined quark delocalization color screening model (QDCSM)~\cite{Pang2004}. In these calculations, binding was attributed to the specific structure of the system, which suppresses the contribution of the color-magnetic interaction. Further studies of the $\Omega N$ dibaryon using the QDCSM and chiral quark models were reported in Refs.~\cite{Zhu2015,Huang2015}. 
The role of nonpairwise three-quark interactions 
has been investigated within perturbative QCD, lattice QCD, 
and phenomenological multiquark models~\cite{Brambilla2010,Takahashi2002,Baek2026}. 
These studies provide additional insight into the microscopic 
dynamics of multiquark systems and the possible contributions 
of many-quark correlations beyond pairwise interactions. 
However, the extent to which such effects influence the effective 
interaction between two color-singlet baryons, particularly
in the $\Omega N$ system, remains an open question. Despite differences in their underlying assumptions and implementations, these studies, together with other theoretical investigations \cite{Goldman1987,Oka1988,Li2000,Pang2004,Zhu2015,Huang2015,SEKIHARA2021}, have consistently suggested the existence of a bound $\Omega N$ state with quantum numbers $J^{\pi}=2^{+}$.

Lattice QCD calculations have provided an independent means of investigating the $\Omega N$ interaction, including studies performed at near the physical quark masses \cite{Etminan2014,Morita2016,LagrangianMethod,Iritani2019}. 
In particular, HAL QCD calculations near the physical quark masses 
revealed a strongly attractive ${}^{5}S_2$ $\Omega N$ potential 
without a pronounced repulsive core and predicted a shallow bound 
state 
\cite{Iritani2019}. 
The relatively large scattering length compared with the effective 
range indicates that the system lies close to unitarity, making 
its near-threshold properties particularly sensitive to the 
interaction strength and coupled-channel effects. In parallel, a formalism for describing scattering involving decuplet baryons has been developed within chiral effective field theory, yielding the minimal Lagrangian and leading-order (SU(3)) potentials for interactions between baryon-octet and baryon-decuplet states \cite{Meissner2017}. This framework has been applied to both $\Omega N$ and $\Omega\Omega$ scattering, comparing results  with lattice QCD calculations. Despite differences in approaches and details among the studies \cite{Goldman1987,Oka1988,Li2000,Pang2004,Zhu2015,Huang2015,SEKIHARA2021,Etminan2014,Morita2016,LagrangianMethod,Iritani2019}, these calculations provide substantial theoretical support for an attractive $\Omega N$ interaction and the possible formation of a near-threshold bound or quasibound state.

Experimental constraints on the $\Omega N$ interaction have also
emerged from femtoscopic measurements of proton--$\Omega$
momentum correlations in relativistic heavy-ion collisions.
The sensitivity of these correlations to the strong interaction underlying
 was theoretically demonstrated~\cite{Morita2016,Morita2020}.
Subsequent measurements by the STAR Collaboration in Au+Au
collisions at RHIC~\cite{STAR2019} and the ALICE Collaboration
at the LHC~\cite{ALICE2020} provided experimental evidence
consistent with a substantial attractive $p\Omega$ interaction.
These results establish an important connection between lattice
QCD predictions, effective interaction models, and experimentally
accessible observables.

The strong attraction predicted in the $\Omega N$ system naturally raises a further question: can this interaction support bound systems containing an $\Omega$ baryon and two or more nucleons? Such few-body systems, including $\Omega NN$, $\Omega NNN$, and $\Omega NNNN$, are particularly intriguing because they combine a baryon carrying three units of negative strangeness with ordinary nuclear constituents. Their study provides an opportunity to investigate how the $\Omega N$ interaction and few-body dynamics cooperate to generate binding in systems beyond the two-body sector. In particular, the pronounced attraction in the ${}^{5}S_2$ 
channel and the sensitivity of the near-threshold $\Omega N$ 
state to the interaction potential motivate a systematic 
investigation of how different short-range $\Omega N$ 
interactions affect the binding energies and spatial structures 
of few-body $\Omega$-nuclear systems. 
Comparing lattice-QCD-derived and meson-exchange potentials 
within the same few-body framework provides a means of assessing 
the model dependence of these predictions.

The binding energy of the lightest $\Omega NN$ system has been investigated in Refs.~\cite{GV0,GV} using momentum-space Faddeev equations \cite{GV2016}. In these calculations, the two-body amplitudes were expanded in Legendre polynomials, with the symmetrization associated with the two identical nucleons taken into account. The $\Omega NN$ system has also been studied using configuration-space Faddeev equations in Refs.~\cite{FilKezPRC2025,FilKezPRD2025,FilKezPRC2026}. In Ref.~\cite{GV0}, the authors employed the local $\Omega N$ potential of Ref.~\cite{LagrangianMethod} to investigate the $\Omega d$ system in the maximal-spin channel, $(I)J^{\pi}=(0)5/2^{+}$. In Ref.~\cite{GV}, the $\Omega NN$ system was studied using the interaction derived by HAL QCD Collaboration \cite{Iritani2019}. The same interaction was subsequently employed in Refs.~\cite{Zhang2022,ESE2023} to calculate the binding energy of the $\Omega NN$ system using different theoretical approaches. 
These studies suggest that the $\Omega NN$ system may be more strongly bound than the $\Omega N$ subsystem. Such an enhancement would reflect the combined effects of the attractive $\Omega N$ interaction and the additional binding generated by the three-body dynamics.

Extending these investigations to systems with four and five constituents is a natural next step. The $\Omega NNN$ and $\Omega NNNN$ systems offer a means of exploring how the binding mechanism evolves with increasing nuclear mass and of testing the predictive power of the underlying baryon--baryon interactions in a genuinely few-body setting. More generally, the existence and properties of bound $\Omega$-nuclear systems could have implications for hypernuclear physics and for our understanding of dense baryonic matter, in which degrees of freedom carrying multiple units of strangeness may become relevant \cite{Tolos2020}.

A related class of systems is formed by $\phi$-mesic nuclei, in which a $\phi$-meson interacts with a nuclear core. The lightest such systems, $\phi NNN$ and $\phi NNNN$, were investigated in Ref.~\cite{LazKezFilPRD2026}. Unlike $\Omega$-nuclear systems, in which strangeness is carried explicitly by a baryon, $\phi$-mesic nuclei contain hidden strangeness in the mesonic component. Although the two classes of systems differ in their constituents and interaction mechanisms, their study provides complementary insight into the role of strange degrees of freedom in few-body nuclear systems. In particular, systems containing an $\Omega$ baryon and three or four nucleons remain largely unexplored.

Another important direction is the investigation of few-body
systems containing two $\Omega$ baryons, corresponding to the
strangeness sector $S=-6$.
The possibility of a bound $\Omega\Omega$ dibaryon was first
explored within quark models~\cite{Zhang2000PRC,Li2001NPA,
Dai2006CPL,Pang2004}, with predictions exhibiting significant
model dependence.
More recently, HAL QCD calculations performed near the physical
quark masses revealed a strongly attractive $\Omega\Omega$
interaction in the ${}^{1}S_{0}$ channel and predicted a
shallow bound state with a binding energy of approximately
$1.6$~MeV, reduced to $0.7$~MeV when the Coulomb repulsion
between the two negatively charged $\Omega$ baryons is
included~\cite{OmegaOmega2018}.
The proximity of the $\Omega\Omega$ system to the unitary
regime motivates extending these investigations to the
$\Omega\Omega N$ and $\Omega\Omega NN$ systems, where the
interplay between the attractive $\Omega N$ and $\Omega\Omega$
interactions, few-body correlations, and Coulomb effects
may lead to the formation of bound states.
A systematic investigation of these systems within the
Faddeev and Faddeev--Yakubovsky frameworks provides an
opportunity to explore the emergence of binding in the
$S=-6$ sector and to assess the sensitivity of the
predicted properties to the underlying baryon--baryon
interactions.

In this work, we present a detailed few-body investigation of $\Omega$-nuclear systems containing up to five baryons. We formulate and solve the Faddeev--Yakubovsky equations in configuration space, providing a unified framework for treating the three-, four-, and five-body systems. The three-body Faddeev equations and the four-body Faddeev--Yakubovsky equations are well-established tools that have been successfully applied to a wide range of quantum systems. In Refs.~\cite{Nogga_Phd,Nogga3NL_02}, the four-body Faddeev--Yakubovsky equations were solved for the first time for a system composed of non-identical particles. Subsequently, Refs.~\cite{Lazauskas2018a,Lazauskas2018b} reported the first solutions of the five-body equations for a system of identical nucleons.

In the present study, we extend this line of research by solving, for the first time, the four- and five-body Faddeev--Yakubovsky equations for $\Omega NNN$, $\Omega NNNN$, and $\Omega \Omega NN$ systems. We employ the HAL QCD $\Omega N$ interactions from Refs.~\cite{LagrangianMethod,Iritani2019}, together with the simplistic Malfliet--Tjon MT I--III ($NN$) interaction~\cite{Malfliet1969,MTcorr}. 
By comparing lattice-QCD-derived and meson-exchange $\Omega N$ 
potentials, we examine the sensitivity of few-body binding to 
the short-range interaction and quantify the effects of Coulomb 
attraction and the successive addition of nucleons on the binding 
energies and spatial structures of the resulting systems. Within this framework, we investigate the binding properties of the $\Omega NNN$, $\Omega NNNN$, and $\Omega \Omega NN$ systems, thereby extending the study of $\Omega$-nuclear binding beyond the three-body sector and providing a systematic assessment of the role of few-body dynamics in these multi-strangeness systems.

The article is organized as follows.
Section~II briefly outlines the configuration-space
Faddeev and Faddeev--Yakubovsky formalism and introduces
the spatial observables used in our analysis.
Section~III describes the adopted $\Omega N$,
$\Omega\Omega$, and $NN$ interactions.
In Sec.~IV, we present and discuss the binding energies
and spatial properties of systems containing a single
$\Omega$ baryon, corresponding to strangeness sector $S=-3$,
including the effects of Coulomb interactions.
Section~V is devoted to the $\Omega\Omega N$ and
$\Omega\Omega NN$ systems with strangeness $S=-6$.
Finally, Sec.~VI summarizes the principal results
and conclusions.

\section{Theoretical framework}

In the present study, we employ the same formalism as in our previous works, based on the solution of the Faddeev--Yakubovsky equations in configuration space. The spin and angular dependence of the wave functions is treated using a partial-wave expansion, while the radial coordinates are discretized using the Lagrange-mesh method. The formalism and its numerical implementation have been described in detail in our earlier publications. In particular, the present approach is fully consistent with that employed in our recent study of $\phi$-mesic nuclei, Refs.~\cite{LazKezFilPRD2026,LazKezFilPLB2026}. We therefore refer the reader to these works for a detailed account of the formalism and numerical procedures, and do not reproduce those details here.

A particular emphasis in the present work is placed on the spatial structure of the few-body systems and, in particular, on the radial properties of their wave functions. We therefore introduce below the relevant definitions of the radii used in our analysis, beginning with the hyperradius and its relation to the Jacobi coordinates employed in this work.

For an $N$-particle system, any independent choice of Jacobi coordinates is associated with a specific partition sequence made up of $(N-1)$ steps, during which the full system is progressively broken down into clusters and subclusters. As a result, a given Jacobi set comprises $(N-1)$ vectors; at each partitioning step $i$, one introduces a vector linking the centers-of-mass of the two clusters, ($a$) and ($b$), formed at that stage:
\begin{equation}
\mathbf{x}_{a,b}^{i}=\sqrt{\frac{2M_{a}M_{b}}{(M_{a}+M_{b})m_{N}}}\left( 
\mathbf{R}_{a}-\mathbf{R}_{b}\right) ,
\end{equation}
where $\mathbf{R}_a$ and $M_a$ denote, respectively, the center-of-mass position vector and the total mass of cluster $a$. The reference mass $m_N$ is introduced to ensure that the Jacobi coordinates have dimensions of length; throughout this work, we choose $m_N$ to be the nucleon mass.

A useful quantity associated with a Jacobi coordinate set is the
hyperradius, defined as the sum of the squared Jacobi coordinates, 
\begin{equation}
\rho ^{2}_N=\sum_{i=1}^{N-1}\left( \mathbf{x}_{a,b}^{i}\right) ^{2}.
\end{equation}

It can equivalently be expressed in terms of the interparticle distances $%
\left\vert \mathbf{r}_{i}-\mathbf{r}_{j}\right\vert $ or particle coordinates
as 
\begin{eqnarray*}
\rho ^{2}_N &=&2\sum_{i<j}^{N}\frac{m_{i}m_{j}}{Mm_{N}}\left| \mathbf{r}_{i}-%
\mathbf{r}_{j}\right| ^{2} \\
&=&2\sum_{i=1}^{N}\frac{m_{i}}{m_{N}}\mathbf{r}_{i}^{2}-2\frac{M}{m_{N}}%
\mathbf{R}_{c.m.}^{2},
\end{eqnarray*}%
where $M=\sum_{i=1}^{N}m_{i}$ is the total mass of the system and 
$\mathbf{R}_{c.m.}=(\sum_{i=1}^{N}m_{i}\mathbf{r}_{i})/{M}
$
is its center-of-mass position.

The latter expression makes explicit that the hyperradius is independent of the choice of Jacobi coordinates and is invariant under translation and rotation of the coordinate system. Thus, although different Jacobi coordinate sets correspond to different clusterization schemes, they all yield the same hyperradius. This invariance makes the hyperradius a particularly useful quantity for characterizing the overall spatial extent of few-body systems, independently of the specific partition used in the Faddeev--Yakubovsky formalism.

\section{Baryon-baryon interactions}
\label{InterPots}

An important element in the present few-body calculations is the simultaneous use of the HAL QCD $\Omega N$ and $\Omega\Omega$ interactions. Both potentials were extracted within mutually consistent lattice-QCD setups. The calculations employed $(2+1)$-flavor  lattice QCD near the physical point, using the Iwasaki gauge action and a nonperturbatively $O(a)$-improved Wilson quark action on a $96^4$ lattice. The lattice spacing was $a\simeq 0.0846$~fm, corresponding to a spatial extent $L\simeq8.1$~fm. Both calculations used the same simulated meson masses, $m_\pi\simeq 146$ MeV and $m_K\simeq 525$ MeV, and yielded an $\Omega$-baryon mass of approximately $1712$ MeV. Thus, the two interactions are consistent with respect to the underlying lattice parameters and hadron masses.
Still the potentials were extracted at different Euclidean-time separations, $t/a=11\text{--}14$ for $\Omega N$ and $t/a=16\text{--}18$ for $\Omega\Omega$, reflecting the different correlation functions and the temporal regions suitable for controlling inelastic and excited-state contamination. Within the respective time windows, the extracted potentials were found to be stable within their uncertainties. Their simultaneous use therefore provides a consistent HAL QCD-based description of the pairwise interactions in the $\Omega\Omega N$ and $\Omega\Omega NN$ systems, subject to the systematic uncertainties associated with the leading-order potential approximation and the use of slightly heavier-than-physical baryon masses.

In the $S$-wave $\Omega N$ system, the two spin channel are possible $^{3}S_1$ and $^{5}S_2$. The $^{3}S_1$ channel is particularly relevant to the study of $\Omega N$ interactions in nuclear medium because it couples strongly to inelastic octet--octet baryon channels at short distances. As demonstrated in lattice QCD studies \cite{Morita2016}, this coupling generates a substantial absorptive component in the effective interaction, opening additional decay channels and potentially reducing the lifetime of an $\Omega N$ state. These effects must be considered when assessing the stability of exotic hypernuclei.

The $\Omega N$ interaction in the strangeness $-3$ sector was investigated in Ref.~\cite{Etminan2014} using the HAL QCD method, in which the potential was extracted from the equal-time Nambu--Bethe--Salpeter wave function calculated in $2+1$-flavor lattice QCD. The initial calculation, performed at a heavy pion mass of $m_\pi=875$ MeV, was a pilot study of this system. Solving the
Schr\"odinger equation with the resulting potential yielded a bound state in the $^{5}S_2$ channel with a binding energy of 18.9 MeV.

Subsequent calculations with near-physical quark masses found a substantially weaker but still attractive interaction in the $^{5}S_2$ channel.
In Ref.~\cite{Iritani2019}, the corresponding HAL QCD potential was used to investigate the $p\Omega^-$ system, including the Coulomb interaction.
The $^{3}S_1$ channel, by contrast, exhibits strong coupling to inelastic channels such as $\Lambda\Xi$ and $\Sigma\Xi$
at separations below approximately 2 fm \cite{Morita2016}. Although this absorption is significant, its effect on the qualitative
behavior of the $\Omega p$ correlation function was found to be limited when compared with the result obtained from $^{5}S_2$ scattering alone \cite{Morita2016}.

In the present study, we focus on the $^{5}S_2$ $\Omega N$ interaction, following Refs.~\cite{GV0,GV,FilKezPRC2025,FilKezPRD2025,FilKezPRC2026,Zhang2022,ESE2023}. This choice allows us to investigate the few-body binding properties associated with the attractive spin-2 channel without explicitly treating the strongly coupled inelastic channels of the $^{3}S_1$ sector.

\textbf{HAL QCD $\Omega N$ potential} as derived in~\cite{Iritani2019} is given:
\begin{equation}
V_{L\Omega N}=b_{1}e^{-b_{2}r^{2}}+b_{3}\left( 1-e^{-b_{4}r^{2}}\right)
\left( \frac{e^{-m_{\pi }r}}{r}\right) ^{2}.  
\label{HALInteraction}
\end{equation}
Here the index ``$L$'' denotes a lattice QCD potential. Four sets of fitting parameters $b_{1}$, $b_{2}$, $b_{3}$, and $b_{4}$ were obtained in Ref.~\cite{Iritani2019}, with $m_{\pi}=146$ MeV and given in Table \ref{Tab:QCDLat}. The resulting scattering characteristics from these parameter sets are consistent within statistical uncertainties. The Yukawa-squared term at large distances is motivated by two-pion exchange between the nucleon and the $\Omega$ baryon.

Four parameter sets were obtained in Ref.~\cite{Iritani2019} for different Euclidean-time separations, $t/a=11\text{--}14$. The corresponding fitted parameters are listed in Table~\ref{Tab:QCDLat}. The resulting scattering observables are consistent among the parameter sets within statistical uncertainties, providing an estimate of the sensitivity to the choice of Euclidean-time separation.
\begin{table}[!htbp]
\centering
\caption{Parameters of the HAL QCD $\Omega N$ interaction
potentials(\ref{HALInteraction}) in physical units at different Euclidean time separations $t/a$ with the statistical errors. Data from Ref. \cite{Iritani2019}.}
\label{Tab:QCDLat}

\begin{tabular}{c c c c c}
\hline\hline
Potential
& $V_{L1\Omega N}$
& $V_{L2\Omega N}$
& $V_{L3\Omega N}$
& $V_{L4\Omega N}$ \\
\hline
$t/a$ & 11 & 12 & 13 & 14 \\
\hline
$b_{1}$ [MeV]
& $-306.5(5.5)$
& $-313.0(5.3)$
& $-316.7(9.4)$
& $-296(18)$ \\

$b_{2}$ [fm$^{-2}$]
& $73.9(4.4)$
& $81.7(5.4)$
& $81.9(8.4)$
& $64(16)$ \\

$b_{3}$ [MeV$\cdot$fm$^{2}$]
& $-266(32)$
& $-252(27)$
& $-237(43)$
& $-272(109)$ \\

$b_{4}$ [fm$^{-2}$]
& $0.78(11)$
& $0.85(10)$
& $0.91(18)$
& $0.76(34)$ \\
\hline\hline
\end{tabular}
\end{table}

\textbf{Meson-exchange $\Omega N$ interaction.} In addition to the HAL QCD interaction, we consider the local $\Omega N$ coordinate-space potential developed 
within a baryon--baryon interaction model based on meson exchange
in Ref.~\cite{LagrangianMethod}. This potential
reads
\begin{equation}
V_{Yu \Omega N}=\frac{\pi^3}{2 r}\sum_{i=1}^{n}C_{n}\left( \frac{\Lambda ^{2}}{%
\Lambda ^{2}-m_{n}^{2}}\right) ^{2}\left[ e^{-m_{n}r}-\frac{\left( \Lambda
^{2}-m_{n}^{2}\right) r+2\Lambda }{2\Lambda }e^{-\Lambda r}\right],
\label{Hyodo}
\end{equation}
where $\Lambda=1$ GeV is the cutoff, $m_n=n\times100$ MeV, and $C_n$ are the strength parameters listed in Table~\ref{tab:Cn_values_row}. The subscript $Yu$ identifies the Yukawa-type interaction.

 The potential was constructed for the $^{5}S_2$ channel and is suitable for few-body calculations. Its long-range component arises from $\eta$-meson exchange and correlated two-meson exchange in the scalar--isoscalar channel, while the short-range component is represented by a contact interaction. Inelastic channels, including $\Lambda\Xi$, $\Sigma\Xi$, and $\Lambda\Xi(1530)$, were incorporated through $K$-meson exchange. Eliminating these channels leads to an energy-dependent effective single-channel interaction, although the resulting energy dependence was found to be small~\cite{LagrangianMethod}.

\begin{table}[!htbp]
\centering
\caption{Values of the parameters $C_n$ in $\text{MeV}\cdot$\text{fm}  
for $\Omega N$ interaction at the fixed $\Lambda =10^3$ MeV and $m_n=n\times 100$ MeV. The coefficients given by the real part of the last column in Table V of Ref. \cite{LagrangianMethod}.}
\label{tab:Cn_values_row}
\begin{tabular}{c c c c c c c c c c}
\hline\hline
$n$   & 1 & 2 & 3 & 4 & 5 & 6 & 7 & 8 & 9 \\
\hline
$C_n$ & $0.14$ & $- 13.76$ & $306.81$ & $- 2729.49$ & $11744.48$ & $- 26288.42$ & $30558.08$ & $- 16730.80$ & $3051.85$ \\
\hline\hline
\end{tabular}
\end{table}

\begin{figure}[!htbp]
\begin{center}
\includegraphics[width=21pc]{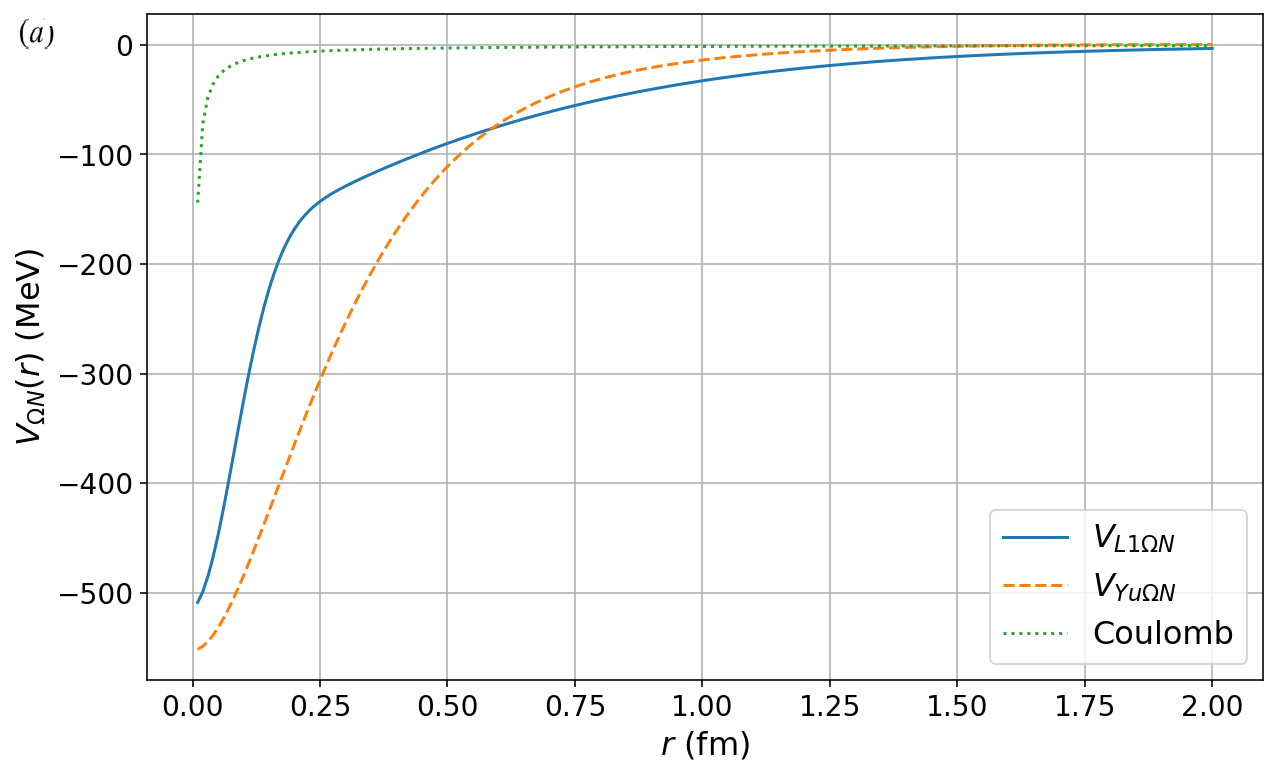}
\includegraphics[width=21pc]{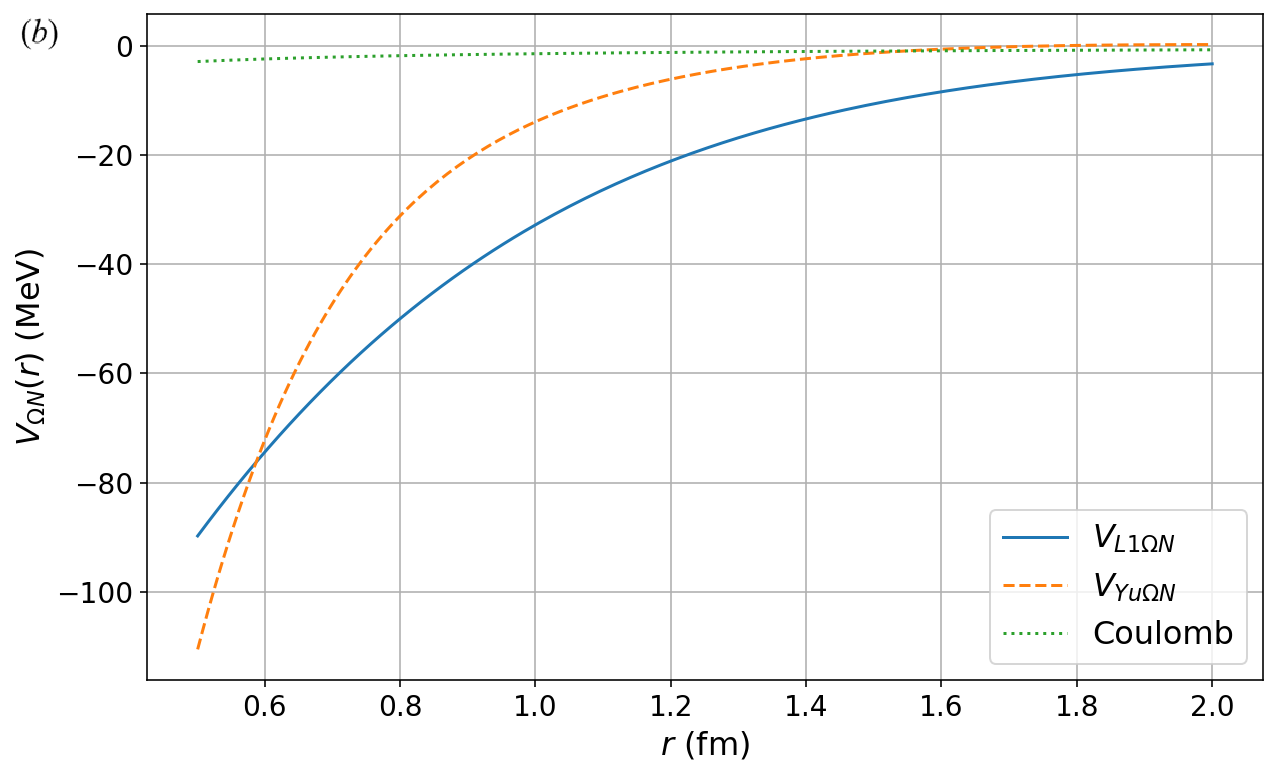}
\end{center}
\caption{Comparison of the Coulomb interaction and the $\Omega^-N$ potentials at short and intermediate distances: (a) $0.01~\mathrm{fm}<r<2~\mathrm{fm}$; (b) $0.5~\mathrm{fm}<r<2~\mathrm{fm}$. Reproduced from Ref. \cite{FilKezPRC2025}.}
\label{fig:41b}
\end{figure}

Figure~\ref{fig:41b} compares the HAL QCD potential in Eq.~(\ref{HALInteraction}) with the meson-exchange potential in Eq.~(\ref{Hyodo}). At short distances, $r<0.6$ fm, the meson-exchange potential is broader and more attractive than the HAL QCD potential. At larger separations, it becomes less attractive than the HAL QCD interaction. Neither potential exhibits the pronounced repulsive core characteristic of realistic nucleon--nucleon interactions. This difference reflects the distinct quark-flavor composition of the $\Omega$-baryon and the nucleon, which suppresses certain quark-level Pauli-blocking effects in the $\Omega N$ system.

Given the substantial uncertainties associated with the available
$\Omega$-baryon interaction models, the present study can provide only a qualitative assessment of the properties of $\Omega$-nuclear systems. Although more accurate nucleon--nucleon interactions could, in principle, be employed, their use would offer limited additional benefit at this stage, since the overall theoretical uncertainties are expected to be dominated by the limited knowledge of the $\Omega N$ and $\Omega\Omega$ interactions. We therefore adopt the simplified MT I--III \cite{Malfliet1969,MTcorr} 
potential, as appropriate to the systems under consideration.

\textbf{{$\Omega\Omega$ interaction.}} 
For the investigation of few-body systems with strangeness
$S=-6$, we employ the $\Omega\Omega$ interaction derived
by the HAL QCD Collaboration from $(2+1)$-flavor lattice
QCD simulations near the physical quark masses
($m_\pi \simeq 146$~MeV)~\cite{OmegaOmega2018}.
The calculations were performed with a lattice spacing
$a \simeq 0.0846$~fm and a spatial volume of
$(8.1~\mathrm{fm})^3$.
In the present study, we consider the $\Omega\Omega$
interaction in the ${}^{1}S_0$ channel, which exhibits
a strong attraction and supports a shallow bound state.
The corresponding local potential is parametrized
as a sum of three Gaussian terms~\cite{OmegaOmega2018}:
\begin{equation}
V_{\Omega\Omega}(r)
=
\sum_{j=1}^{3}
c_j
\exp\left[-\left(\frac{r}{d_j}\right)^2\right].
\label{OmegaOmega}
\end{equation}
\begin{table}[!htbp]
\centering
\caption{Parameters of the
$\Omega\Omega$ interaction potential in
Eq.~(\ref{OmegaOmega}) at Euclidean time
$t/a=17$~\cite{OmegaOmega2018}.}
\begin{tabular}{ccc}
\hline\hline
$j$ & $c_j$ (MeV) & $d_j$ (fm) \\
\hline
1 & $914(52)$  & $0.143(5)$ \\
2 & $305(44)$  & $0.305(29)$ \\
3 & $-112(13)$ & $0.949(58)$ \\
\hline\hline
\end{tabular}
\end{table}

The resulting $\Omega\Omega$ potential is characterized
by short-range repulsion and intermediate-range attraction,
with the attractive component sufficiently strong to
support a weakly bound dibaryon. The Pauli principle restricts the allowed $S$-wave states for the $\Omega\Omega$ system to $^{1}S_{0}$ and $^{5}S_{2}$. The results of the HAL QCD Collaboration \cite{QCD2015,OmegaOmega2018} indicate that the $^{1}S_{0}$ channel exhibits a strongly attractive interaction. 
Unlike the nucleon--nucleon interaction, the
$\Omega\Omega$ interaction does not contain a long-range
one-pion-exchange contribution.
In the calculations of the $\Omega\Omega N$ and
$\Omega\Omega NN$ systems, this potential is combined
with the $\Omega N$ and $NN$ interactions described
above, allowing us to investigate the interplay of
pairwise interactions and few-body correlations in
the $S=-6$ sector.

\section{$\Omega$ systems with strangeness $-3$}

In this section, we investigate few-body systems containing
a single $\Omega$ baryon and two, three, or four nucleons,
corresponding to the strangeness sector $S=-3$.
Using the configuration-space Faddeev and
Faddeev--Yakubovsky equations, we analyze their binding
energies and spatial structures for different $\Omega N$
interaction potentials, with particular emphasis on the
effects of Coulomb attraction and the successive addition
of nucleons.

\subsection{$\Omega N$ system}

Table~\ref{tab:ON} summarizes the calculated ground-state properties of the 
two-body $\Omega N$ system in the $J^\pi=2^+$ state, obtained using four HAL QCD
$\Omega N$ potentials. For the $\Omega^- n$ system, in which only the strong 
interaction is included, the binding energies range from 1.2160 to 1.3762 MeV, while the 
corresponding rms interparticle distances, $d$, lie between 3.991 and 4.202 fm. 
The variation among the four potentials is approximately 5\%. Including the attractive 
Coulomb interaction in the $\Omega^- p$ system enhances the binding, yielding binding 
energies of $2.0831-2.2648$ MeV and reducing the rms distances to $3.390-3.514$ fm, with a 
variation of approximately 4\%. The Coulomb contribution to the binding energy, $\Delta E_C$, is remarkably stable across the four potentials, ranging from 0.867 to 0.889 MeV, 
corresponding to a variation of only about 2.5\%. The relatively weak dependence of these 
observables on the choice of the $V_{L1\Omega N}-V_{L4\Omega N}$ 
potentials reflects their calibration to reproduce the $\Omega N$ binding energy 
predicted by lattice QCD.

For comparison, the meson-exchange potential of Ref.~\cite{LagrangianMethod}, when used together with its imaginary optical component, predicts the $\Omega^- n$ system to be a resonant state located only 36 keV above threshold, while the $\Omega^- p$ system is bound by 865 keV. The substantial imaginary component of this potential accounts for the instability of the $\Omega$ baryon and its enhanced decay in the presence of a nucleon. By absorbing probability flux from the elastic channel, this absorptive term reduces the effective attraction in the $\Omega N$ system. Consequently, removing the imaginary component of the potential allows the system to develop stronger binding. Remarkably,
Coulomb attraction still provides similar difference of binding energies as HAL QCD potentials.
Thus, Coulomb attraction contributes approximately 0.88 MeV of additional binding and results in a more compact $\Omega^- p$ system compared to $\Omega^- n$ compound.


\begin{table}[!htbp]
\centering
\caption{
Two-body $\Omega N$ system in the $J^\pi=2^+$ state. 
The results for the ground state energy for the $\Omega^- n$ system include only the strong interaction, whereas 
the $\Omega^-p$ results include both the strong and Coulomb
interactions. The Coulomb contribution to the binding energy is 
defined as $\Delta E_C=|E_{\Omega^-p}-E_{\Omega^- n}|$. $r_{xy}$ is  root mean square (rms) distance between tbe
particles $x$ and $y$.}

\label{tab:ON}

\begin{tabular}{lccccc}

\hline\hline

 & \multicolumn{2}{c}{$\Omega^- n$}

 & \multicolumn{2}{c}{$\Omega^-p$}

 & $\Delta E_C$ \\
 \hline


Potential

& $E_2$ [MeV] &  $r_{\Omega n}$ 
[fm]

& $E_2$ [MeV] & $r_{\Omega p}$ [fm]

& [MeV] \\

\hline

$V_{L1\Omega N}$ & $-1.291$ & 4.111 & $-2.160$ & 3.472 & 0.869 \\

$V_{L2\Omega N}$ & $-1.376$ & 3.991 & $-2.265$ & 3.390 & 0.889 \\

$V_{L3\Omega N}$ & $-1.216$ & 4.203 & $-2.083$ & 3.514 & 0.867 \\

$V_{L4\Omega N}$ & $-1.366$ & 4.014 & $-2.246$ & 3.412 & 0.880 \\
$V_{Yu\Omega N}$ & $0.036-0.691i$& $3.62-3.46i$  &  $-0.865-1.030i$ & $3.617-1.678i$ & $0.901-0.339i$\\

Re($V_{Yu\Omega N}$) & $-0.275$ & 7.963 & $-1.082 $ & 4.541 & 0.807 \\
\hline\hline

\end{tabular}

\end{table}


\FloatBarrier
\subsection{$\Omega NN$ system}

We extend the few-body analysis beyond the two-body $\Omega N$ subsystem by considering the three-body $\Omega NN$ system with total isospin $T=0$.
Since the $\Omega$ baryon has isospin $I_\Omega=0$, the two nucleons form an isosinglet $np$ pair, and the system therefore corresponds to $\Omega^-np$ compound.
Table~\ref{tab:ONN} summarizes the calculated properties of the $\Omega^-np$ system for the $J^\pi=5/2^+$ and $3/2^+$ states obtained with five considered $\Omega N$ potentials, both with and without the Coulomb interaction. 
\begin{table}[!htbp]

\centering

\small

\caption{
Three-body $\Omega NN$ system with $T=0$, calculated without
and with the Coulomb interaction. Here, $\rho^2_3$ is
the mean-square hyperradius, defined as the sum of the squares of the
mass-scaled Jacobi coordinates of the three-body system.
The quantities $r_\Omega^2$ and $r_N^2$ denote the mean-square distances
of the $\Omega$-baryon and nucleons, respectively, from the center-of-mass of
the $\Omega np$ system, and $\langle T_3\rangle$ is the expectation
value of the kinetic energy of the three-body system.}

\label{tab:ONN}

\begin{tabular}{llccccc}

\hline\hline

Potential & Coulomb
& $E_3$ [MeV]
& $\rho^2_3$ [fm$^2$]
& $\langle T_3\rangle$ [MeV]
& $ r_\Omega^2$ [fm$^2$]
& $ r_N^2$ [fm$^2$] \\

\hline

\multicolumn{7}{c}{$\Omega np$, \qquad $J^\pi=5/2^+$} \\

\hline

$V_{L1\Omega N}$ & No  & -19.572 & 8.22 & 57.7 & 0.605 & 1.52 \\
                 & Yes & -20.882 & 8.03 & 58.8 & 0.587 & 1.48 \\

$V_{L2\Omega N}$ & No  & -20.000 & 8.07 & 58.8 & 0.591 & 1.49 \\
                 & Yes & -21.323 & 7.88 & 59.9 & 0.574 & 1.46 \\

$V_{L3\Omega N}$ & No  & -19.421& 8.17 & 58.4 & 0.600 & 1.51 \\
                 & Yes & -20.740 & 7.97 & 59.6 & 0.582 & 1.47 \\

$V_{L4\Omega N}$ & No  & -19.854 & 8.16 & 58.0 & 0.600 & 1.51 \\
                 & Yes & -21.167 & 7.97 & 59.1 & 0.582 & 1.47 \\

Re($V_{Yu\Omega N}$) & No  & -16.339  & 7.587 & 76.3 & 0.545 & 1.41 \\
                 & Yes & -17.876  & 7.33 & 78.4 & 0.521 & 1.37  \\

$V_{Yu\Omega N}$i & No  & --- & --- & --- & --- & --- \\
     & Yes & $(-17.754,-4.114)$
           & $(7.17,-0.842)$
           & $(78.8,5.55)$
           & $(0.505,-0.0746)$
           & $(1.34,-0.144)$ \\

\hline

\multicolumn{7}{c}{$\Omega np$, \qquad $J^\pi=3/2^+$} \\

\hline

$V_{L1\Omega N}$ & No  & -3.004 & 59.1 & 18.0 & 6.60 & 8.00 \\
                 & Yes & -3.598 & 46.2 & 19.6 & 4.55 & 7.21 \\

$V_{L2\Omega N}$ & No  & -3.061 & 57.0 & 18.4 & 6.25 & 7.90 \\
                 & Yes & -3.666 & 45.1 & 20.0 & 4.38 & 7.13 \\

$V_{L3\Omega N}$ & No  & -2.952 & 61.2 & 17.9 & 6.95 & 8.09 \\
                 & Yes & -3.541 & 47.0 & 19.6 & 4.67 & 7.27 \\

$V_{L4\Omega N}$ & No  & -3.055 & 57.3 & 18.2 & 6.29 & 7.91 \\
                 & Yes & -3.657 & 45.4 & 19.8 & 4.41 & 7.14 \\

Re($V_{Yu\Omega N}$)  & No  & -2.329 & 253. & 15.5 & 35.2 & 23.4 \\
                      & Yes & -2.801 & 77.7 & 19.7 & 9.32 & 10.5 \\

$V_{Yu\Omega N}$i & No  & --- & --- & --- & --- & --- \\
     & Yes & $(-2.623,-0.567)$
           & $(44.4,-38.5)$
           & $(19.6,6.03)$
           & $(4.44,-5.76)$
           & $(7.40,-4.06)$ \\

\hline\hline

\end{tabular}

\end{table}

Within our considered models $\Omega$ baryon interacts with nucleon in $J^\pi=2^+$ channel only, therefore the three-body state with a  total angular momentum 
$J^\pi=5/2^+$ is the most favorable as it preserves all three interparticle interactions active at the same time. Still $\Omega^-np$ compound can be bound with total
angular momentum $J^\pi=3/2^+$. On the other hand, $J^\pi=1/2^+$ states can not be bound, since in this configuration $\Omega$-baryon does not interact with nucleons in the $J^\pi=2^+$ channel.

As in calculations for a two-body system we asses the model 
dependence presenting results for four parametrizations of Lattice QCD interaction as well as using meson-exchange potential of Ref.~\cite{LagrangianMethod}.
Coulomb interaction provides additional attraction between the negatively charged $\Omega^-$ and the proton. 
By keeping the strong interactions unchanged, we can isolate the effects of Coulomb attraction on the binding energies and spatial structure. 
The resulting eigenenergies, kinetic energies, and characteristic spatial observables for both states are also reported in Table~\ref{tab:ONN}.

For the $J^\pi=5/2^+$ state, Coulomb attraction increases 
the binding energy by  $1.309$--$1.323$~MeV, with only a weak dependence on the choice of $\Omega N$ potential.
The system also becomes more compact, as reflected in the decrease of $\rho^2_3$ from $\rho^2_3$ from
$8.071$--$8.222$~fm$^2$ to $7.883$--$8.027$~fm$^2$. 
Naturally, the $J^\pi=3/2^+$ state is substantially more weakly bound and 
spatially more extended. Without Coulomb attraction, its energy ranges from $-2.952$ to $-3.061$~MeV and $\rho^2_3$ from
$56.95$ to $61.21$~fm$^2$. Inclusion of the Coulomb interaction
increases the binding by $0.589$--$0.605$~MeV and produces a pronounced
contraction of the system, reducing $\rho^2_3$ to
$45.14$--$46.96$~fm$^2$. This contraction is also reflected in the
substantial decrease of the $\Omega$'s mean-square distance $r_\Omega^2$,
whereas the nucleon mean-square distance $r_N^2$ changes more
moderately. As a system becomes stronger bound $\Omega$-baryon tries to establish in the center between the nucleons, stimulated by its larger mass (smaller kinetic energy) and nucleon-baryon interaction being of shorter range than nucleon-nucleon one.

In Table \ref{tab:T1_3B} we summarize our calculations of the $\Omega NN$ systems with total 
isospin $T=1$. One can have three isospin configurations reflecting $\Omega^-nn$, $\Omega^-np$, and
$\Omega^-pp$ particle content. Bound $T=1$ states exist only with a total angular momenta $J^\pi=3/2^+$,
which allows T=1 $NN$ pair to exploit favorable $^1S_0$ configuration. This system, being stronger 
constrained by Pauli principle between nucleons,
is even less favorable than $T=0$ one with the same angular momentum. The $\Omega^-pp$  is the strongest 
bound due to presence of two attractive Coulomb interaction between proton and $\Omega^-$ baryon, whereas $\Omega^-nn$
system is weakest bound. In this system $\Omega$ baryon settles inbetwenn the nucleons -- therefore average distance
between the nucleons is considerably larger than between the $\Omega$-baryon and nucleon. This feature also reflects increase of binding energy 
from  $\Omega^-np$ to $\Omega^-pp$ system by almost 400 keV, which is result of competition between two attractive $\Omega-p$ interactions and
a repulsive $p-p$ one. Notably, if one compares the binding energies of the $\Omega^-nn$ and $\Omega^-np$ systems, one can estimate the 
weight of the Coulomb interaction of $\Omega-p$ to $\approx$ 600 keV.

\begin{table}[!htbp]
\centering
\caption{
Binding energies and average kinetic energies, in MeV, for three-body $\Omega NN$ systems with a total isospin $T=1$. 
}

\label{tab:T1_3B}

\begin{tabular}{lcccccc}

\hline\hline

 & \multicolumn{2}{c}{$\Omega^- nn$}

 & \multicolumn{2}{c}{$\Omega^-np$}

 & \multicolumn{2}{c}{$\Omega^-pp$}
 
\\
 \hline


Potential

& $E_3$  &  $\langle T_3\rangle$  & $E_3$  & $\langle T_3\rangle$  & $E_3$  & $\langle T_3\rangle$ 

 \\

\hline

$V_{L1\Omega N}$ & $-2.241$ & $18.9$ & $-2.851$ & 20.3 & -3.239 & 21.1 \\

$V_{L2\Omega N}$ & $-2.350$ & $19.5$ & $-2.971$ & 20.9 & -3.367 & 21.7 \\

Re($V_{Yu\Omega N}$) & $-0.727$ & $19.0$ & $-1.311$  & 22.5 & -1.735 & 24.7\\

$V_{Yu\Omega N}$ & $-0.484-1.078i$ & $20.7-9.14i$  &  $-1.108-1.232i$ & $23.65-8.08i$ & $-1.553-1.336i$ & $25.68-7.52i$\\

\hline\hline

\end{tabular}

\end{table}

\begin{table}[!htbp]
\centering
\caption{
Root mean square radii for three-body $\Omega NN$ systems with a total isospin $T=1$ presented in units of fm$^2$. 
}

\label{tab:T1_r3B}

\begin{tabular}{lccccccccc}

\hline\hline

 & \multicolumn{3}{c}{$\Omega^- nn$}

 & \multicolumn{3}{c}{$\Omega^-np$}

 & \multicolumn{3}{c}{$\Omega^-pp$}
 
\\
 \hline


Potential

& $ \rho^2_3$  &  $ r^2_\Omega$ &  $ r^2_N$ 
& $ \rho^2_3$  &  $ r^2_\Omega$ &  $ r^2_N$
& $\rho^2_3$  &  $ r^2_\Omega$ &  $ r^2_N$

 \\

\hline

$V_{L1\Omega N}$ & 54.1 & 3.37 & 10.4 & 47.9 & 2.96  & 9.27 & 47.4 & 2.84 & 9.23\\

$V_{L1\Omega N}$ & 52.6 & 3.27 & 10.2 & 46.8 & 2.88  & 9.06 & 46.3 & 2.76 & 9.03\\

Re($V_{Yu\Omega N}$)  & 103. & 6.73 & 19.6 & 46.8 & 4.76  & 14.5 & 70.5 & 4.17 & 13.8\\

$V_{Yu\Omega N}$ & 59.0  & 3.42  & 11.6  & 56.4  & 3.32   & 11.0  & 56.7 &  3.23  & 11.2\\
          & -47.2$i$ & -3.19$i$ & -8.76$i$ & -29.1$i$ & -2.10$i$  & -5.44$i$ & -24.8$i$ & -1.66$i$ & -4.74$i$\\

\hline\hline

\end{tabular}

\end{table}

We do not observe any bound $\Omega NN$ systems with negative parity or angular momenta different from aforementioned $J=3/2^+$ and $J=5/2^+$ configurations.
This is direct consequence of S-wave dominant interaction models, which favors $L=0^+$ total orbital momentum states.

\FloatBarrier
\subsection{$\Omega NNN$ system}

Having established the existence and properties of the bound
$\Omega NN$ states, we now consider the four-body  $\Omega NNN$ system to determine 
how an additional nucleon affects its binding and spatial distribution.
For total isospin $T=1/2$, the system has two mirror configurations,  $\Omega^-nnp$ (${}^{4}_{\Omega}\mathrm{H}$) 
and $\Omega^-ppn$ (${}^{4}_{\Omega}\mathrm{He}$). In the absence of Coulomb interactions, these mirror systems are degenerate within the charge-independent strong-interaction model used here. We first examine this limit and then include the Coulomb interaction to quantify the resulting differences between the two systems.

Table~\ref{tab:ONNN} summarizes the calculated properties of the  $T=1/2$ four-body
 system without including Coulomb
interactions. As expected from charge independence, 
${}^{4}_{\Omega}\mathrm{H}$ ($\Omega^-nnp$) and
${}^{4}_{\Omega}\mathrm{He}$ ($\Omega^-ppn$) have identical energies and spatial observables in this limit.  All interaction models considered support bound states with total angular momenta $J^\pi=2^+$ and $1^+$, whereas the $J^\pi=3^+$ and $0^+$ remain unbound. This behavior is associated  with the restrictions imposed by the Pauli principle, which renders 
three-nucleon subsystem with $J^\pi=3/2^+$ repulsive. 
Likewise for three-body system -- all four HAL QCD interactions yield almost identical results: therefore here we present results only for its $L1$ and $L2$ models together with the meson-exchange model. The latter produces somewhat weaker binding but preserves the same qualitative pattern.

 The $J^\pi=2^+$ state is  substantially  more deeply bound than the $J^\pi=1^+$ state.
 The calculated energies range from $E_4 \approx-27.1$ to $-31.8$~MeV for $J^\pi=2^+$, compared with  
 $E_4 \approx-11.7$ to $-14.0$~MeV for $J^\pi=1^+$. This difference  is reflected in the spatial distributions: the mean-square hyperradius is about  $\rho^2_4\simeq14-15$~fm$^2$ for a compact
$J^\pi=2^+$ state and
 $\rho^2_4\simeq25.$--$28.$~fm$^2$. for the
$J^\pi=1^+$ state. 
The corresponding kinetic-energy expectation values are approximately
$110$~MeV and $65$~MeV, respectively, consistent with the more compact structure of the more deeply bound $J^\pi=2^+$ state. The relatively modest variation across the interaction models suggests that these qualitative features do not dependent on the particular $\Omega N$ interaction model.

Another important aspect is that, owing to the pronounced spin dependence of the $\Omega N$ interaction, the nucleon distribution within the $\Omega NNN$ compound system is highly anisotropic, in contrast to that of a bound three-nucleon system. Nucleons whose spins are aligned with the baryon tend to localize closer to the center, whereas nucleons with opposite spin orientation reside at larger distances. This behavior is particularly evident in the more strongly bound $2^+$ state, for which $R_{N1}^2 < R_{N2}^2$, indicating that the mean-squared distance of the spin-aligned nucleon from the center-of-mass is substantially smaller than the corresponding average for the two nucleons coupled to total spin zero.

We next examine the Coulomb interaction, which lifts the degeneracy between the mirror systems because they have different charge compositions. The results are listed in  Table~\ref{tab:ONNNC}. 
For $J^\pi=2^+$, Coulomb attraction increases the binding of ${}^{4}_{\Omega}\mathrm{H}$ by
$\sim2.5-3.0$~MeV. The increase is even larger for ${}^{4}_{\Omega}\mathrm{He}$ ranging $\sim3.5-4.0$~MeV. The same trend holds for the $J^\pi=1^+$ state. The same tendency is found for  $J^\pi=1^+$:
the binding increases by  $\sim1.4$ MeV in ${}^{4}_{\Omega}\mathrm{H}$ and $\sim2.9$ MeV in
${}^{4}_{\Omega}\mathrm{He}$.

The larger Coulomb shift in
${}^{4}_{\Omega}\mathrm{He}$  is consistent with the spatial organization found in the three-body systems. The heavy $\Omega^-$ baryon is not subject to nucleonic Pauli blocking and tends to occupy the central region, while the nucleons are distributed around it. Consequently, the $\Omega^-$-nucleon separations are, on average, smaller than the separations between nucleons, making the attractive $\Omega^-$–proton Coulomb contributions more effective
than repulsive contribution from $p-p$ pair. The Coulomb interaction also contracts the four-body wave functions. This contraction is modest for the already compact $2^+$ state but more pronounced for the spatially extended $1^+$ one. Thus, Coulomb effects both split the mirror-state energies and modify their spatial structure; the size of each effect depending on the state’s initial spatial distribution.

\begin{table}[!htbp]

\centering

\small

\caption{Four-body $\Omega NNN$ system with $T=1/2$, corresponding to
${}^{4}_{\Omega}\mathrm{H}$ and ${}^{4}_{\Omega}\mathrm{He}$ (here $^4_{\Omega}\mathrm{H}$ and
$^4_\Omega\mathrm{He}$ nuclei, where the superscript $A$ is a total number of baryons), calculated
without the Coulomb interaction. The $J^\pi=3^+$ and $0^+$ states are
unbound. Here, $\rho^2_4$ is the mean-square hyperradius,
defined as the sum of the squares of the three mass-scaled Jacobi
coordinates of the four-body system. The quantities $r_\Omega^2$,
$r_{N2}^2$, and $r_{N1}^2$ denote the mean-square distances of the
$\Omega$ baryon, the pair of identical nucleons, and the remaining
nucleon, respectively, from the center-of-mass of the $\Omega NNN$
system. The quantity $\langle T_4\rangle$ is the expectation value of
the kinetic energy of the four-body system.}
\label{tab:ONNN}
\begin{tabular}{lcccccc}

\hline\hline

Potential

& $E_4$ [MeV]

& $\langle T_4\rangle$ [MeV]

& $\rho^2_4$ [fm$^2$]

& $ r^2_\Omega$ [fm$^2$]

& $ r^2_{N2}$ [fm$^2$]

& $ r^2_{N1}$ [fm$^2$] \\

\hline
\multicolumn{7}{c}{$J^\pi=2^+$} \\

\hline
$V_{L1\Omega N}$ & -31.382 & 109. & 14.7 & 0.776 & 2.21 & 1.53 \\

$V_{L2\Omega N}$ & -31.840 & 111. & 15.2 & 0.763 & 2.20 & 1.51 \\


Re($V_{Yu\Omega N}$)  & -27.085 & 124. & 14.4 & 0.727 & 2.22 & 1.46 \\

$V_{Yu\Omega N}$i & $(-26.982,-4.426)$ & $(123.,9.0)$ & $(14.3,-0.85)$
     &  &  &  \\

\hline

\multicolumn{7}{c}{$J^\pi=1^+$} \\
\hline
$V_{L1\Omega N}$ & -13.910 & 64.2 & 25.0 & 1.94  & 2.87 & 3.18 \\

$V_{L2\Omega N}$ & -14.052 & 64.7 & 24.8   & 1.90  & 2.85 & 3.17 \\


Re($V_{Yu\Omega N}$)  & -11.702 & 66.3 & 28.2 & 2.44 & 3.10 & 3.37 \\

$V_{Yu\Omega N}$i & $(-11.571,-1.585)$ & $(66.6,6.8)$ & $(26.8,-4.46)$
     & --- & --- & --- \\

\hline

\end{tabular}
\end{table}

Therefore, the Coulomb interaction has two related effects on the
four-body $\Omega NNN$ systems: it lifts the degeneracy between the
mirror ${}^{4}_{\Omega}\mathrm{H}$ and
${}^{4}_{\Omega}\mathrm{He}$ states and simultaneously increases
their binding and reduces their spatial extension. The magnitude of
these effects depends not only on the number of protons but also on
the spatial structure of the corresponding four-body state.

\subsection{$\Omega NNNN$ system}

Having established the binding and structural properties of the
four-body $\Omega NNN$ systems, we proceed to the next member of the
few-body hierarchy by adding one more nucleon. The resulting five-body
$\Omega NNNN$ system is of particular interest because the four
nucleons can form a tightly bound ${}^{4}\mathrm{He}$ ($\alpha$-particle)
core. Thus, the system can be viewed as an $\Omega^-$ baryon coupled
to an $\alpha$ particle, $\Omega^-+\alpha$, corresponding to
${}^{5}_{\Omega}\mathrm{He}$ compound.  In contrast to the  $T=1/2$ mirror pair encountered in the four-body system, the configuration considered here has total isospin $T = 0$, with two protons and two neutrons. We calculate the $J^\pi=3/2^+$ state using the all five different $\Omega N$ interaction models and examine the effect of the Coulomb interaction.\footnote{Our five body code does not allow treating optical potentials, wherefore analysis for meson-exchange $\Omega N$ potential is limited to calculation using its real part only.}

The calculated energies are summarized in Table~\ref{tab:O4N}.
Without Coulomb interactions, the four HAL QCD parametrizations 
give energies between $-59.8$ and $-60.6$ MeV.
The total spread is 0.74 MeV, or only about 1.2\% of the binding energy.
Less attractive meson-exchange model  provide by $\sim 6$ MeV less binding. 
 

Including the Coulomb interaction preserves the same structure and ordering of the predicted energies,
simply shifting all the energies downward by $\sim1.6$ MeV. This suggests that the different models yield similar average sizes and spatial distributions of the particles in this five-body system.

Interestingly, this Coulomb shift is almost a factor of two smaller than that found for the ${}^4_\Omega\mathrm{He}$ state, despite the two systems having the same charge content. This observation suggests that the more strongly bound ${}^4\mathrm{He}$ core is less susceptible to structural rearrangement induced by the presence of the $\Omega^-$ baryon. Through its spin-dependent interaction, the $\Omega^-$ tends to attract nucleons with aligned spins while repelling those with opposite spin orientations, thereby modifying the internal structure of the nuclear core. This effect is more readily manifested when the $\Omega^-$ is introduced into the less strongly bound, although still relatively compact, ${}^3\mathrm{He}$ system. In contrast, the stronger binding of the ${}^4\mathrm{He}$ core makes it considerably more resistant to such rearrangements, highlighting its greater structural rigidity.



\begin{table}[!htbp]
\centering
\small
\caption{
Four-body $\Omega^-nnp$ and $\Omega^-ppn$ systems,
${}^{4}_{\Omega}\mathrm{H}$ and ${}^{4}_{\Omega}\mathrm{He}$,
respectively, with $T=1/2$, where the superscript $A=4$ denotes the
total baryon number, calculated with the Coulomb interaction. Here, $\rho^2_4$ is the mean-square
hyperradius, defined as the sum of the squares of the three mass-scaled
Jacobi coordinates of the four-body system. The quantities
$r_\Omega^2$, $r_{n}^2$, and $r_{p}^2$ denote the mean-square
distances from the systems center-of-mass for the $\Omega$ baryon, neutron and proton respectively.}
\label{tab:ONNNC}
\begin{tabular}{llcccccc}
\hline\hline
System & Potential

& $E_4$ [MeV]

& $\langle T_4\rangle$ [MeV]

& $\rho^2_4$ [fm$^2$]

& $r_\Omega^2$ [fm$^2$]

& $r_{n}^2$ [fm$^2$]

& $r_{p}^2$ [fm$^2$] \\

\hline

\multicolumn{8}{c}{$J^\pi=2^+$} \\

\hline

$_\Omega^4\mathrm{H}$

& $V_{L1\Omega N}$ & -33.99 & 114. & 14.4 & 0.75  & 2.20 & 1.48 \\

& $V_{L2\Omega N}$ & -34.38 & 115. & 14.3 & 0.74 & 2.18 & 1.46 \\

& $V_{Yu\Omega N}$  & -29.99 & 129. & 14.04 & 0.690 & 2.20 & 1.38 \\

& $V_{Yu\Omega N}$i & $(-29.89,-4.52)$ & $(129.,8.42)$
                    & $(13.95,-0.751)$ & --- & --- & --- \\

$_\Omega^4\mathrm{He}$

& $V_{L1\Omega N}$ & -34.88 & 116. & 14.1 & 0.73  & 1.50 & 2.13\\

& $V_{L2\Omega N}$ & -35.37 & 117. & 14.0 & 0.72  & 1.49 & 2.12\\

& $V_{Yu\Omega N}$ & -31.03 & 137. & 13.7 & 0.67  & 1.42 & 2.12\\

& $V_{Yu\Omega N}$i & $(-30.91,-4.54)$ & $(131.,8.63)$
       & $(13.6,-0.72)$ & --- & --- & --- \\

\hline

\multicolumn{8}{c}{$J^\pi=1^+$} \\

\hline

$_\Omega^4\mathrm{H}$

& $V_{L1\Omega N}$ & -15.36 & 67.8 & 23.6 & 1.76 & 2.77 & 3.00 \\

& $V_{L2\Omega N}$ & -15.50 & 68.3 & 23.3 & 1.73 & 2.75 & 2.99 \\

& $V_{Yu\Omega N}$  & -13.08 & 69.6 & 25.6 & 2.08 & 2.93 & 3.14 \\

& $V_{Yu\Omega N}$i & $(-12.97,-1.67)$ & $(69.9,6.44)$
       & $(24.8,-3.22)$ & --- & --- & --- \\

$_\Omega^4\mathrm{He}$

& $V_{L1\Omega N}$ & -16.74 & 70.5 & 22.6 & 1.60  & 3.03  &2.66\\

& $V_{L2\Omega N}$ & -16.92 & 71.1 & 22.4 & 1.57  & 3.02 & 2.64\\

& $V_{Yu\Omega N}$  & -14.76 & 75.91 & 23.41 & 1.71  & 3.12 & 2.73\\

& $V_{Yu\Omega N}$i & $(-14.66,-1.87)$ & $(75.4,6.28)$
       & $(23.0,-2.08)$ & --- & --- & --- \\

\hline

\end{tabular}

\end{table}

\begin{table}[!htbp]
\centering
\caption{Five-body $\Omega NNNN$, in state $J^\pi=3/2^+$, $T=0$.}
\label{tab:O4N}
\begin{tabular}{lcc}
\hline\hline
Potential & \multicolumn{2}{c}{$E$ [MeV]} \\
\hline
          & no Coulomb & with Coulomb \\
\hline
$V_{L1}$ & -60.0 & -61.6 \\
$V_{L2}$ & -60.6 & -62.1 \\
$V_{L3}$ & -59.8 & -61.3 \\
$V_{L4}$ & -60.4 & -61.9 \\
$V_{Yu}$ & -54.5 & -56.2 \\
\hline

\end{tabular}
\end{table}

\section{$\Omega$ systems with strangeness $-6$}

The $\Omega\Omega$ dibaryon, with total strangeness $S=-6$, is an especially interesting system in the high-strangeness sector. HAL QCD lattice calculations indicate that the $\Omega\Omega$ interaction in the ${}^{1}S_{0}$ channel  is  overall attractive and supports a shallow bound state with a binding energy of order  $\sim 1$ MeV. The large scattering length and relatively small effective range suggest that the system lies close to the unitary regime.

The structure of the $\Omega\Omega$ interaction differs considerably from that of the nucleon--nucleon interaction. In particular, quark-level Pauli-blocking effects are reduced, and the absence of pion exchange makes the attractive interaction comparatively short-ranged. The $\Omega\Omega$ system therefore offers a useful setting in which to examine baryon–baryon dynamics dominated by short-range QCD effects. Lattice-QCD calculations provide a direct link between the underlying interaction and the resulting scattering and binding observables. The large scattering length may also enhance low-relative-momentum correlations, making $\Omega\Omega$ system potentially accessible in relativistic heavy-ion collision experiments.

The possible shallow $\Omega\Omega$  bound state motivates the study of systems containing additional nucleons, such as $\Omega\Omega N$ and $\Omega\Omega NN$. These systems provide a framework for examining how near-unitary two-body dynamics influence binding in larger few-body systems and whether correlations between binding energies emerge that are relatively insensitive to short-distance details. They also allow the role of three- and higher-body correlations to be investigated using interactions constrained by lattice QCD. In the present work, we focus on the $\Omega\Omega N$  and $\Omega\Omega NN$  systems, applying configuration-space Faddeev and Faddeev–Yakubovsky methods.

\subsection{$\Omega\Omega N$ system}

The $\Omega\Omega N$ system is a useful probe of baryon–baryon interactions involving heavy strange quarks. Its structure is governed by the competition between the attractive $\Omega N$ interaction and the short-range behavior of the $\Omega\Omega $ interaction. We solve the configuration-space Faddeev equations using HAL QCD interactions for both subsystems. We also combine the HAL QCD $\Omega\Omega$ potential with the meson-exchange $\Omega N$ interaction to assess the sensitivity of the three-body predictions to the choice of $\Omega N$ force.

Binding energy of $\Omega\Omega N$ states has been calculated by solving the Faddeev equations in momentum space \cite{GV0}, using the hyperspherical harmonic approach for Faddeev equations \cite{ESE2023}, the variational method \cite{Zhang2022}, and within the framework of Faddeev equations in configuration space \cite{FilKezPRC2026}.

Our calculations are performed using both the baryon masses associated with the lattice-QCD interactions and the physical masses, as specified for each interaction combination. The results for the $J^\pi=1/2^+$ and $5/2^+$ states are presented in Tables \ref{tab:OON12} and \ref{tab:OON52}, respectively. The tables report the energies, kinetic-energy expectation values, mean-square hyperradii, and mean-square distances of the baryons from the center-of-mass.

For $J^\pi = 1/2^+$ state represent the most favorable configuration, in which all three
particle pairs interact.
The four HAL QCD $\Omega N$ potentials predict a bound $\Omega\Omega N$ state in the absence of Coulomb interactions, with energies between $-5.9$ and $-6.2$
MeV. The small spread, about $0.28$ MeV, indicates limited sensitivity to the differences among these parametrizations. The corresponding mean-square hyperradii lie between $17.81$ and $18.30$ fm$^2$, while the mean-square distances associated with the $\Omega^-$ baryons and the nucleon remain close to $3.5$ and $1.5$ fm$^2$, respectively. The meson-exchange $\Omega N$ potential also predicts a bound state, but with weaker binding ($E_3=-4.291$ MeV) and a somewhat larger mean-square hyperradius of $20.3$ fm$^2$. Thus, both interaction models support a bound and relatively compact $J^\pi = 1/2^+$ state, although the predicted binding depends moderately on the  interaction model.

Coulomb effects differ between $\Omega^-\Omega^- n$ and $\Omega^-\Omega^- p$. In $\Omega\Omega n$, the Coulomb contribution is repulsive between the two negatively 
charged $\Omega^-$ baryons; there is no proton to provide an attractive $\Omega^-$-proton term. The binding is therefore reduced. For example, with $V_{L1}$ the
energy changes from $-6.0$ to $-4.9$ MeV, and with $V_{L2}$ from $-6.2$ to $-5.1$ MeV. The mean-square hyperradius increases accordingly, indicating a more 
diffuse state.

In $\Omega^-\Omega^- p$, two attractive $\Omega^-$–proton Coulomb interactions compete with 
the repulsive $\Omega^-\Omega^-$ interaction. 
The net Coulomb contribution is attractive: with the HAL QCD potentials, the energies range from approximately
$-6.83$ to $-7.12$ MeV, compared with $-5.90$ to $-6.18$ MeV without Coulomb. The mean-square 
hyperradius decreases to about $16.6-17.0$ fm$^2$, showing that the system becomes more compact.
The meson-exchange calculation exhibits the same qualitative trend.

In $J^\pi = 5/2^+$ state two $\Omega$ baryons can not exploit favorable $^1S_0$ configuration. Therefore this state is more weakly bound and substantially
more extended. Without Coulomb interactions, the HAL QCD potentials give energies between $-2.22$ and $-2.50$ MeV and mean-square hyperradii 
of approximately $55-60$ fm$^2$, compared with about  
$18$ fm$^2$ for the $J^\pi = 1/2^+$  state.
The meson-exchange potential produces an even more weakly bound state, with $E_3=-0.5653 $ 
MeV and $ \rho^2_3  = 172.3$ fm$^2$. These results indicate that the $5/2^+$ state is
much more sensitive to the adopted $\Omega N$  interaction.

The extended structure of the $5/2^+$ state also makes it particularly sensitive to Coulomb effects. 
In  $\Omega\Omega n$, the repulsive Coulomb intteraction between two baryons   reduces the binding predicted by the HAL QCD 
potentials by approximately $1.8$ MeV in magnitude and increases $\rho^2_3$ to about $85$ fm$^2$. With the meson-exchange interaction, 
the energy moves from $-0.565$ to $-0.292$ MeV and the mean-square hyperradius increases to $1030$ fm$^2$, indicating a state extremely close to the breakup threshold. In $\Omega^-\Omega^- p$, the net Coulomb interaction instead enhances binding and contracts the system. For the HAL QCD potentials, the energies
range from $-3.31$ to $-3.61$ MeV and $\rho^2_3$ decreases to about $50$ fm$^2$. With the meson-exchange potential, 
the energy changes to $-1.673$ MeV and $\rho^2_3$ decreases to $\sim88$  fm$^2$.

In summary, we predict $J^\pi = 1/2^+$  and $5/2^+$  $\Omega\Omega N$  bound states.
The $J^\pi = 3/2^+$ configuration does not 
support a bound state -- since in this state 
two baryons can not interact in favorable $^1S_0$ state, while also only one interacting $\Omega N$ pair can be formed in interacting $2^+$ configuration. Negative parity states are also not
supported within used fremework, based on dominant $S$-wave interactions, which mostly support total orbital momenta $L=0^+$ bound states


\begin{table}[!htbp]
\centering
\small

\caption{
Three-body $\Omega\Omega N$ system in the $J^\pi=1/2^+$ state,
calculated without and with the Coulomb interaction. The Coulomb
calculations are presented separately for the $\Omega\Omega n$ and
$\Omega\Omega p$ systems. Here, $\rho^2_3$ is the
mean-square hyperradius, defined as the sum of the squares of the
mass-scaled Jacobi coordinates of the three-body system.
The quantities $r_\Omega^2$ and $r_N^2$ denote the mean-square
distances of an $\Omega$ baryon and the nucleon, respectively, from
the center-of-mass of the $\Omega\Omega N$ system, and
$\langle T_3\rangle$ is the expectation value of the kinetic energy
of the three-body system.}

\label{tab:OON12}

\begin{tabular}{llccccc}
\hline\hline
System & Potential
& $E_3$ [MeV]
& $\langle T_3\rangle$ [MeV]
& $\rho^2_3$ [fm$^2$]
& $r_N^2$ [fm$^2$]
& $r_\Omega^2$ [fm$^2$] \\
\hline

\multicolumn{7}{c}{$J^\pi=1/2^+$} \\
\hline

$\Omega\Omega N$ & $V_{L1\Omega N}$ & -6.001 & 33.3 & 18.2 & 3.60 & 1.54 \\
(no Coulomb)     & $V_{L2\Omega N}$ & -6.178 & 34.1 & 17.8 & 3.48 & 1.52 \\
                 & $V_{L3\Omega N}$ & -5.896 & 33.6 & 18.3 & 3.62 & 1.55 \\
                 & $V_{L4\Omega N}$ & -6.135 & 33.6 & 18.0 & 3.53 & 1.52 \\
                 & $V_{Yu\Omega N}$ & -4.291 & 41.4 & 20.3 & 4.28 & 1.64 \\
                 & $V_{Yu\Omega N}$i & $(-4.113,-1.827)$
                         & $(42.4,7.70)$
                         & $(17.7,-5.68)$
                         & $(3.42,-1.66)$
                         & $(1.52,-0.333)$ \\

\hline

$\Omega\Omega n$ & $V_{L1\Omega N}$ & -4.94 & 31.8 & 19.9 & 3.78 & 1.72 \\
(Coulomb)        & $V_{L2\Omega N}$ & -5.11 & 32.6 & 19.4 & 3.65 & 1.69 \\
                 & $V_{L3\Omega N}$ & -4.83 & 32.1 & 20.0 & 3.80 & 1.73 \\
                 & $V_{L4\Omega N}$ & -5.07 & 32.2 & 19.6 & 3.70 & 1.70 \\
                 & $V_{Yu\Omega N}$ & -3.23 & 39.3 & 22.7 & 4.62 & 1.88 \\
                 & $V_{Yu\Omega N}$i & $(-3.0421,-1.7731)$
                         & $(40.37,8.034)$
                         & $(19.44,-6.827)$
                         & $(3.601,-1.869)$
                         & $(1.710,-0.4344)$ \\

\hline

$\Omega\Omega p$ & $V_{L1\Omega N}$ & -6.925 & 35.1 & 17.0 & 3.12 & 1.50 \\
(Coulomb)        & $V_{L2\Omega N}$ & -7.125 & 36.0 & 16.6 & 3.03 & 1.48 \\
                 & $V_{L3\Omega N}$ & -6.828 & 35.5 & 17.0 & 3.12 & 1.50 \\
                 & $V_{L4\Omega N}$ & -7.069 & 35.4 & 16.8 & 3.12 & 1.50 \\
                 & $V_{Yu\Omega N}$  & -5.424 & 46.2 & 17.2 & 3.21 & 1.51 \\
                 & $V_{Yu\Omega N}$i & $(-5.283,-2.028)$
                         & $(46.9,-6.66)$
                         & $(16.0,-3.67)$
                         & $(2.86,-0.949)$
                         & $(1.44,-0.249)$ \\

\hline\hline
\end{tabular}
\end{table}

\begin{table}[!htbp]
\centering
\small

\caption{
Three-body $\Omega\Omega N$ system in the $J^\pi=5/2^+$ state,
calculated without and with the Coulomb interaction. The Coulomb
calculations are presented separately for the $\Omega\Omega n$ and
$\Omega\Omega p$ systems. The quantities $\rho^2_3$,
$r_\Omega^2$, $r_N^2$, and $\langle T_3\rangle$ have the same
definitions as in Table~\ref{tab:OON12}.}

\label{tab:OON52}

\begin{tabular}{llccccc}
\hline\hline
System & Potential
& $E_3$ [MeV]
& $\langle T_3\rangle$ [MeV]
& $\rho^2_3$ [fm$^2$]
& $r_\Omega^2$ [fm$^2$]
& $r_N^2$ [fm$^2$] \\
\hline

\multicolumn{7}{c}{$J^\pi=5/2^+$} \\
\hline

$\Omega\Omega N$ & $V_{L1\Omega N}$ & -2.350 & 20.0 & 58.1 & 6.26 & 6.30 \\
(no Coulomb)     & $V_{L2\Omega N}$ & -2.502 & 21.0 & 54.9 & 5.91 & 5.95 \\
                 & $V_{L3\Omega N}$ & -2.222 & 20.0 & 60.4 & 6.47 & 6.55 \\
                 & $V_{L4\Omega N}$ & -2.481 & 20.5 & 55.6 & 6.00 & 6.03 \\
                 & $V_{Yu\Omega N}$  & -0.565 & 19.6 & 172. & 17.8 & 18.9\\
                 & $V_{Yu\Omega N}$i & $(-0.047,-1.34)$
                         & $(23.4,17.3)$
                         & $(16.9,-75.08)$
                         & $(1.80,-7.60)$
                         & $(1.80,-8.21)$ \\

\hline

$\Omega\Omega n$ & $V_{L1\Omega N}$ & -1.848 & 17.97 & 84.3 & 8.20 & 9.26 \\
(Coulomb)        & $V_{L2\Omega N}$ & -1.984 & 18.92 & 78.0 & 7.64 & 8.57 \\
                 & $V_{L3\Omega N}$ & -1.728 & 17.88 & 89.8 & 8.65 & 9.88 \\
                 & $V_{L4\Omega N}$ & -1.968 & 18.51 & 79.1 & 7.75 & 8.69 \\
                 & $V_{Yu\Omega N}$  & -0.292 & 12.94 & 1028 & 72.7 & 117. \\
                 & $V_{Yu\Omega N}$i & $(0.335,-1.030)$
                         & $(18.7,18.8)$
                         & $(-15.2,-98.8)$
                         & $(-1.06,-9.77)$
                         & $(-1.85,-10.9)$ \\

\hline

$\Omega\Omega p$ & $V_{L1\Omega N}$  & -3.433 & 22.3 & 49.6 & 5.21 & 5.41 \\
(Coulomb)        & $V_{L2\Omega N}$  & -3.609 & 23.3 & 47.1 & 4.95 & 5.14 \\
                 & $V_{L3\Omega N}$  & -3.307 & 22.5 & 50.9 & 5.32 & 5.56 \\
                 & $V_{L4\Omega N}$  & -3.576 & 22.8 & 47.8 & 5.03 & 5.21 \\
                 & $V_{Yu\Omega N}$  & -1.673 & 27.9 & 88.1 & 8.61 & 9.73 \\
                 & $V_{Yu\Omega N}$i & $(-1.293,-1.733)$
                         & $(29.8,14.0)$
                         & $(36.8,-50.8)$
                         & $(3.84,-4.88)$
                         & $(4.01,-5.62)$ \\

\hline\hline
\end{tabular}
\end{table}

\subsection{$\Omega\Omega NN$ system}

The calculated properties of the four-body $\Omega\Omega NN$  systems are summarized in Tables~\ref{tab:OONNT1} and \ref{tab:OONNT0}. We first discuss the $T=1$ states in Tables~\ref{tab:OONNT1}. Without Coulomb interactions, both $J^\pi=0^+$ and $2^+$ states are bound for all real $\Omega N$ interactions considered, but their binding energies and spatial structures differ substantially. For the HAL QCD potentials $V_{L1}$ and $V_{L2}$, the $0^+$ energies are $-19.2$ and $-19.7$ MeV, respectively, whereas the $2^+$ energies are $-6.9$ and $-7.2$ MeV. This difference is reflected in the mean-square hyperradius: approximately $15.5$ fm$^2$ for the $0^+$ state, compared with 35–36 fm$^2$ for the $2^+$ state. The $0^+$ state is therefore considerably more compact.

The meson-exchange $\Omega N$ interaction gives the same qualitative ordering but weaker binding: $E_4 = -14.862$ MeV for $J^\pi=0^+$ and $-3.241$ MeV for $J^\pi=2^+$. The latter state is also more extended, with $\rho^2_4$=47.68 fm$^2$. Thus, the $2^+$ state is more sensitive to the choice of $\Omega N$ interaction than the more deeply bound $0^+$ state.

Coulomb effects depend strongly on the charge composition. In $\Omega^-\Omega^- nn$,
the only Coulomb contribution is the repulsion between the two $\Omega^-$ baryons, which reduces the binding. 
With $V_{L1}$, for example, the energy changes from -19.22 to -17.012 MeV 
for $J^\pi=0^+$ and from -6.909 to -6.078 MeV for $J^\pi=2^+$. 
The mean-square hyperradius increases, particularly for the $2^+$ state, 
consistent with a more diffuse configuration.

In $\Omega^-\Omega^- pp$, the repulsive $\Omega^-\Omega^-$ and proton-proton
interactions compete with four attractive $\Omega^-$–proton pair 
interactions. The net Coulomb contribution is attractive. With $V_{L1}$, the 
energies shift from -19.22 to -24.903 MeV for $J^\Pi=0^+$ and from -6.909 to 
-12.198 MeV for $J^\pi=2^+$. The corresponding mean-square hyperradii 
decrease from 15.72 to 14.33 fm$^2$ and from 35.84 to 27.88 fm$^2$, 
respectively, indicating Coulomb-induced contraction. The same qualitative 
behavior is obtained with $VL2$ and the meson-exchange potential.

For $\Omega\Omega np$, two attractive $\Omega^-$–proton interactions compete 
with the repulsive $\Omega^-\Omega^-$ interaction. The net Coulomb 
contribution remains attractive but is smaller than in $\Omega\Omega pp$. 
With $V_{L1}$, the $0^+$ energy changes from -19.22 to -21.113 MeV, while the 
$2^+$ energy changes from -6.909 to -9.212 MeV. For a fixed strong 
interaction, the energies therefore follow the ordering
\[
E_4(\Omega^-\Omega^- pp)<E_4(\Omega^-\Omega^- np)
<E_4(\Omega^-\Omega^- NN)<E_4(\Omega^-\Omega^-nn),
\]
where a more negative energy corresponds to stronger binding. This ordering reflects the balance between attractive $\Omega^-$–proton and repulsive $\Omega^- - \Omega^-$ and proton–proton Coulomb terms.

The $T = 0$ results in Table~\ref{tab:OONNT0} show a different ordering of
the angular-momentum states. Both $J^\pi=1^+$ and $3^+$ are deeply bound and 
compact, but the $3^+$  state is more deeply bound. Without Coulomb 
interactions, $V_{L1}$ gives energies of -31.092 MeV for $1^+$ and -33.916 MeV 
for $3^+$; $VL2$ gives -31.687 and -34.784 MeV, respectively. The meson-exchange potential produces the same ordering, with energies of -27.409 and -30.204 MeV.

The spatial observables are consistent with this pattern. For the HAL QCD interactions, $\rho^2_4$ is approximately 12.4–12.6 fm$^2$ for the $1^+$ state and 11.7–12.0 fm$^2$ for the $3^+$ state. The nucleon mean-square distances are also smaller in the $3^+$ state, indicating a somewhat more compact configuration. This difference is more pronounced with the meson-exchange potential, for which $\rho^2_4$ decreases from 11.46 fm$^2$ in the $1^+$ state to 9.703 fm$^2$ in the $3^+$ state.

The Coulomb interaction further increases the binding of the $T=0$
$\Omega\Omega np$ system. With $V_{L1}$, the energies change from -31.092 to -33.611 MeV for $J^\pi=1^+$  and from -33.916 to -37.540 MeV for $J^\pi=3^+$, corresponding to additional binding of approximately 2.52 and 3.62 MeV. Similar shifts are obtained with $V_{L2}$: about 2.55 MeV for  $1^+$ and 3.67 MeV for  $3^+$. The larger Coulomb shift in the  $3^+$ state is consistent with its more compact spatial distribution.

Coulomb effects also break proton–neutron symmetry in the spatial distributions. Without Coulomb interactions, the proton and neutron mean-square distances are equal. In $\Omega^-\Omega^- np$, the proton is attracted to both $\Omega^-$ baryons and consequently has a smaller mean-square distance from the center-of-mass than the neutron. With $V_{L1}$, for example, the proton and neutron values are 1.46 and 1.54 fm$^2$, respectively, for $J^\pi=1^+$, and 1.25 and 1.30 fm$^2$ for $J^\pi=3^+$. The same trend is found with the other interactions.

Overall, the four-body calculations predict bound $\Omega\Omega NN$ states for all real strong-interaction models considered. The $T=0$ states are more deeply bound and more compact than the $T=1$ states. Within $T=1$, the $0^+$ state is more deeply bound and more compact than the $2^+$ state; within $T=0$, the $3^+$ state is more deeply bound and slightly more compact than the $1^+$ state. Coulomb interactions produce substantial charge-dependent shifts: they reduce binding in $\Omega^-\Omega^- nn$ and increase it in $\Omega^-\Omega^- np$ and, most strongly, $\Omega^-\Omega^- pp$.

For the $T=1$ $\Omega\Omega NN$ system, the $J^\pi=2^+$
state is substantially more extended than the $0^+$ state,
with $\rho_4^2$ increasing from approximately
$15.5$ to $35$--$36$~fm$^2$ (hyperradii of $3.9$ and
$6.0$~fm, respectively).
This reflects its weaker binding and different spin
recoupling, which modifies the contributions of the
attractive $\Omega N({}^{5}S_2)$ and
$\Omega\Omega({}^{1}S_0)$ configurations.

\begin{table}[!htbp]
\centering
\footnotesize

\caption{
Four-body $\Omega\Omega NN$ systems with $T=1$, calculated without
and with the Coulomb interaction. Results are presented for the
$J^\pi=0^+$ and $2^+$ states. Here, $\rho^2_4$ denotes
the mean-square four-body hyperradius, while
$r^2_\Omega$ and $r^2_N$ denote
the corresponding mean-square spatial quantities for the $\Omega$
baryons and nucleons, respectively. The quantity $\langle T_4\rangle$
is the expectation value of the kinetic energy of the four-body
system.}

\label{tab:OONNT1}

\resizebox{\textwidth}{!}{%
\begin{tabular}{lllccccc}
\hline\hline
System & $J^\pi$ & Potential
& $E_4$ [MeV]
& $\langle T_4\rangle$ [MeV]
& $\rho^2_4$ [fm$^2$]
& $r^2_\Omega$ [fm$^2$]
& $r^2_N$ [fm$^2$] \\
\hline

$\Omega\Omega NN$
& $0^+$ & $V_{L1\Omega N}$ & -19.22  & 89.2 & 15.7 & 0.953 & 2.23 \\
(no Coulomb)
&       & $V_{L2\Omega N}$ & -19.66 & 90.5  & 15.5 & 0.938 & 2.19 \\
&       & $V_{Yu\Omega N}$ & -14.86 & 104. & 15.5 & 0.929 & 2.22 \\
&       & $V_{Yu\Omega N}$i& $(-14.69,-4.43)$ & $(105.,8.08)$
               & $(15.1,-2.09)$ & --- & --- \\

& $2^+$ & $V_{L1\Omega N}$ & -6.909 & 49.39 & 35.84 & 2.62 & 4.09 \\
&       & $V_{L2\Omega N}$ & -7.193 & 50.67 & 34.85 & 2.54 & 4.00 \\
&       & $V_{Yu\Omega N}$  & -3.241 & 54.58 & 47.68 & 3.58 & 5.37 \\
&       & $V_{Yu\Omega N}$i & $(-2.857,-3.025)$ & $(56.8,17.7)$
               & $(37.1,-18.0)$ & --- & --- \\

\hline

$\Omega\Omega pp$
& $0^+$ & $V_{L1\Omega N}$ & -24.903 & 101. & 14.3 & 0.889 & 1.99 \\
(Coulomb)
&       & $V_{L2\Omega N}$ & -25.403 & 102. & 14.1 & 0.877 & 1.96 \\
&       & $V_{Yu\Omega N}$  & -21.478 & 121. & 13.5 & 0.834 & 1.88 \\
&       & $V_{Yu\Omega N}$i & $(-21.342,-4.778)$ & $(122.,11.9)$
               & $(13.3,-1.37)$ & --- & --- \\

& $2^+$ & $V_{L1\Omega N}$ & -12.198 & 62.88 & 27.88 & 2.01 & 3.23 \\
&       & $V_{L2\Omega N}$ & -12.556 & 64.22 & 27.25 & 1.96 & 3.17 \\
&       & $V_{Yu\Omega N}$  & -9.048  & 78.52 & 29.41 & 2.12 & 5.62 \\
&       & $V_{Yu\Omega N}$i & $(-8.802,-3.731)$ & $(79.62,13.78)$
               & $(27.34,-6.76)$ & --- & --- \\

\hline

$\Omega\Omega nn$
& $0^+$ & $V_{L1\Omega N}$ & -17.012  & 85.2 & 16.3 & 1.01  & 2.27 \\
(Coulomb)
&       & $V_{L2\Omega N}$ & -17.436  & 86.5 & 16.0 & 0.990 & 2.23 \\
&       & $V_{Yu\Omega N}$  & -12.615 & 98.5 & 16.3 & 0.993 & 2.29 \\
&       & $V_{Yu\Omega N}$i & $(-12.434,-4.361)$ & $(99.2,10.4)$
               & $(15.75,-2.32)$ & --- & --- \\

& $2^+$ & $V_{L1\Omega N}$ & -6.078 & 47.1 & 38.6 & 2.85 & 4.32 \\
&       & $V_{L2\Omega N}$ & -6.350 & 48.4 & 37.5 & 2.76 & 4.20 \\
&       & $V_{Yu\Omega N}$  & -2.520 & 50.8 & 55.6 & 4.26 & 6.04 \\
&       & $V_{Yu\Omega N}$i & $(-2.101,-2.843)$ & $(53.0,18.5)$
               & $(39.7,-22.8)$ & --- & --- \\

\hline

$\Omega\Omega np$
& $0^+$ & $V_{L1\Omega N}$ & -21.113 & 93.22 & 15.2 & 0.940 & 2.11 \\
(Coulomb)
&       & $V_{L2\Omega N}$ & -21.577 & 94.54 & 14.9 & 0.926 & 2.08 \\
&       & $V_{Yu\Omega N}$  & -17.176 & 110.48 & 14.6 & 0.900 & 2.04 \\
&       & $V_{Yu\Omega N}$i & $(-17.021,-4.594)$ & $(111.,10.5)$
               & $(14.3,-1.72)$ & --- & --- \\

& $2^+$ & $V_{L1\Omega N}$  & -9.212 & 55.2 & 31.9 & 2.33 & 3.63 \\
&       & $V_{L2\Omega N}$  & -9.528 & 56.9 & 31.1 & 2.26 & 3.55 \\
&       & $V_{Yu\Omega N}$  & -5.757 & 65.4 & 37.1 & 2.74 & 4.21 \\
&       & $V_{Yu\Omega N}$i & $(-5.450,-3.361)$ & $(67.0,15.5)$
               & $(32.52,-11.11)$ & --- & --- \\

\hline\hline
\end{tabular}%
}
\end{table}

\begin{table}[!htbp]
\centering
\small

\caption{
Four-body $\Omega\Omega NN$ system with $T=0$, calculated without
and with the Coulomb interaction. The Coulomb calculation corresponds
to the $\Omega\Omega np$ system. Results are presented for the
$J^\pi=1^+$ and $3^+$ states. Here,
$\rho^2_4$ denotes the mean-square four-body hyperradius,
and $r^2_\Omega$ denotes the corresponding
mean-square spatial quantity for the $\Omega$ baryons. The proton and neutron
quantities, $r^2_p$ and
$r^2_n$, are given separately. The quantity
$\langle T_4\rangle$ is the expectation value of the kinetic energy
of the four-body system.}

\label{tab:OONNT0}

\resizebox{\textwidth}{!}{%
\begin{tabular}{lllcccccc}
\hline\hline
System & $J^\pi$ & Potential
& $E_4$ [MeV]
& $\langle T_4\rangle$ [MeV]
& $\rho^2_F$ [fm$^2$]
& $r^2_\Omega$ [fm$^2$]
& $r^2_p$ [fm$^2$]
& $r^2_n$ [fm$^2$] \\
\hline

$\Omega\Omega NN$
& $1^+$ & $V_{L1\Omega N}$ & -31.092 & 118. & 12.6 & 0.884 & 1.56 & 1.56 \\
(no Coulomb)
&       & $V_{L2\Omega N}$ & -31.687 & 119. & 12.4 & 0.872 & 1.53 & 1.53 \\
&       & $V_{Yu\Omega N}$  & -27.409 & 140. & 11.5 & 0.815 & 1.40 & 1.40 \\
&       & $V_{Yu\Omega N}$i & $(-27.295,-5.366)$ & $(142.4,13.1)$
               & $(11.33,-0.931)$ & --- & --- & --- \\

& $3^+$ & $V_{L1\Omega N}$ & -33.916 & 126. & 12.0 & 0.936 & 1.32 & 1.32 \\
&       & $V_{L2\Omega N}$ & -34.784 & 128. & 11.7 & 0.911 & 1.29 & 1.29 \\
&       & $V_{Yu\Omega N}$  & -30.204 & 170. & 9.70 & 0.749 & 1.08 & 1.08 \\
&       & $V_{Yu\Omega N}$i & $(-30.003,-7.836)$ & $(171.0,16.5)$
               & $(9.439,-1.248)$ & --- & --- & --- \\

\hline

$\Omega\Omega np$
& $1^+$ & $V_{L1\Omega N}$ & -33.611 & 122. & 12.31 & 0.877 & 1.46 & 1.54 \\
(Coulomb)
&       & $V_{L2\Omega N}$  & -34.236 & 123. & 12.13 & 0.865 & 1.44 & 1.51 \\
&       & $V_{Yu\Omega N}$  & -30.495 & 149. & 11.10 & 0.801 & 1.29 & 1.38 \\
&       & $V_{Yu\Omega N}$i & $(-30.391,-5.471)$ & $(146.,10.7)$
               & $(11.0,-0.832)$ & --- & --- & --- \\

& $3^+$ & $V_{L1\Omega N}$  & -37.540 & 134. & 11.5 & 0.897 & 1.25  & 1.30 \\
&       & $V_{L2\Omega N}$  & -38.449 & 136. & 11.3 & 0.874 & 1.22  & 1.27 \\
&       & $V_{Yu\Omega N}$  & -34.739 & 177. & 9.17 & 0.703 & 0.995 & 1.04 \\
&       & $V_{Yu\Omega N}$i & $(-34.558,-7.963)$ & $(177.,15.7)$
               & $(8.97,-1.06)$ & --- & --- & --- \\

\hline\hline
\end{tabular}%
}
\end{table}

\FloatBarrier

\begin{figure}[H]
\centering
\includegraphics[width=0.50\columnwidth]{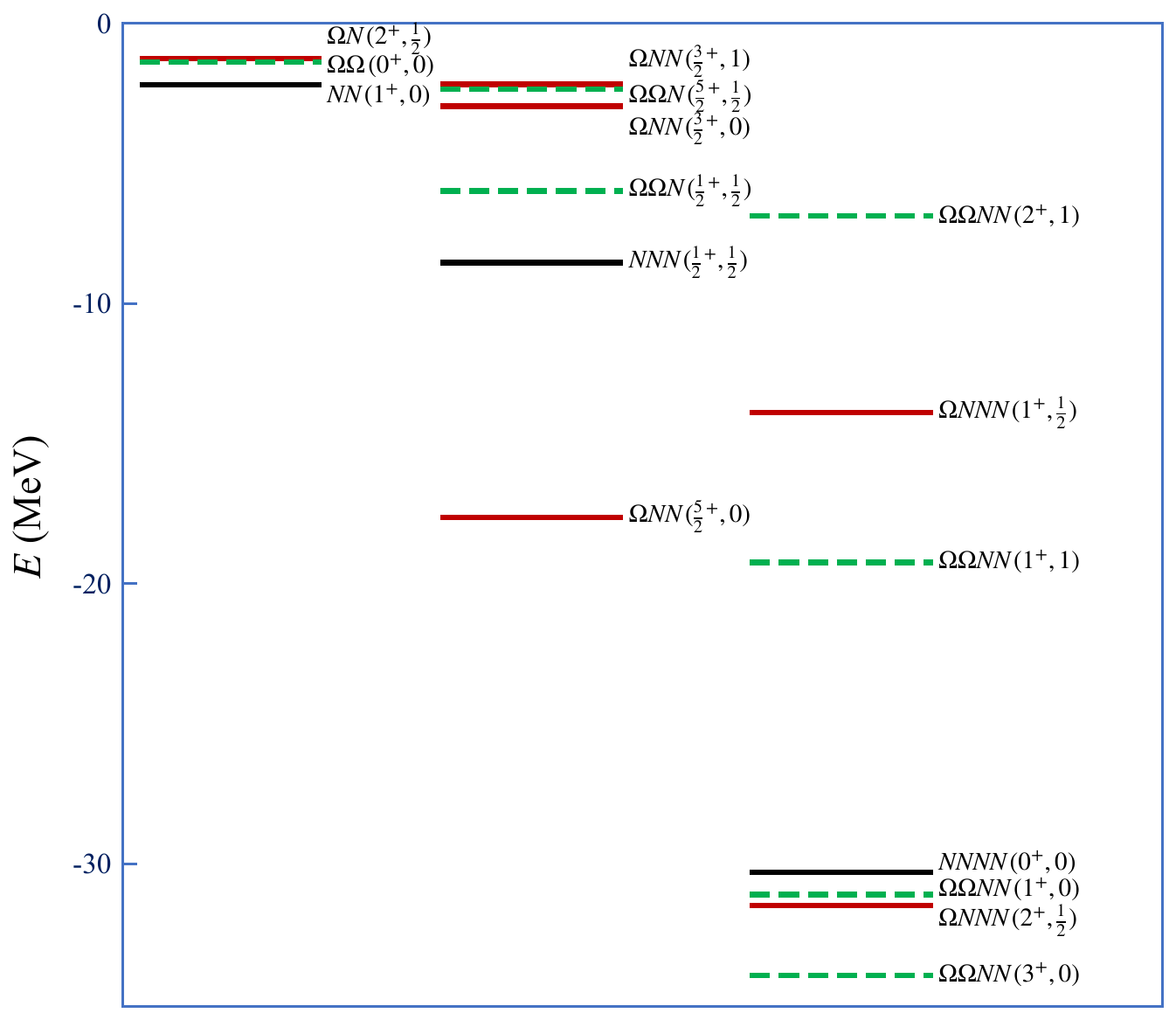}
\caption{Schematic energy spectrum of $\Omega$-nucleonic 
bound states calculated in the isospin approximation, 
neglecting Coulomb interactions and using the 
$V_{L1\Omega N}$ potential. 
Solid lines represent states with strangeness $S=-3$ 
and ordinary nuclear systems, while dashed lines 
represent states with strangeness $S=-6$.}
\label{spectrum}
\end{figure}

Figure~\ref{spectrum} summarizes the calculated energy
spectrum of few-body systems with strangeness $S=-3$ and
$S=-6$ in the absence of Coulomb interactions.
The spectrum illustrates the substantial enhancement of
binding in selected multibaryon configurations and the
pronounced dependence of the binding energies on the
spin--isospin quantum numbers.

\FloatBarrier
\section{Summary and Concluding Remarks}

In this study, we carried out a systematic analysis of the binding energies and spatial characteristics of few-body systems that include one or two $\Omega$ baryons, associated with the strangeness sectors $S=-3$ and $S=-6$. The computations were performed using configuration-space Faddeev and Faddeev–Yakubovsky approaches, combining lattice-QCD-inspired HAL QCD and meson-exchange $\Omega N$ interactions with the HAL QCD $\Omega\Omega$ potential and the MT I–III nucleon–nucleon force. For the first time, we present microscopic calculations for the four- and five-body $\Omega NNN$, $\Omega NNNN$, and $\Omega\Omega NN$ systems, thereby pushing studies of $\Omega$-nuclear binding beyond the three-body cases explored previously.

In the $S=-3$ sector, we predict that $\Omega NN$, $\Omega NNN$, and $\Omega NNNN$ form bound states whose binding strengths and sizes depend strongly on the chosen spin channels. Notably, the $J^\pi=2^+$ four-body $\Omega NNN$ states are much more deeply bound and more compact than the corresponding $1^+$ states. The five-body $\Omega NNNN$ system with $J^\pi=3/2^+$ is found to be significantly bound, with energies around -60 MeV for HAL QCD-based interactions when Coulomb effects are neglected. In the $S=-6$ sector, we predict bound $\Omega\Omega N$ states with $J^\pi=1/2^+$ and $5/2^+$, along with bound $\Omega\Omega NN$ states with $J^\pi=0^+,2^+$ for $T=1$ and $J^\pi=1^+,3^+$ for $T=0$. Among these, the four-body $T=0$, $3^+$ state stands out as especially strongly bound and compact, underscoring the role of spin-dependent $\Omega N$ and $\Omega\Omega$ correlations.

An important feature of the predicted spectrum is its
pronounced dependence on underlying spin--isospin configurations.
In the $S=-3$ sector, the $J^\pi=5/2^+$ state of
$\Omega NN$ and the $J^\pi=2^+$ state of $\Omega NNN$
are considerably more strongly bound than their
lower-spin counterparts.
Similarly, in the $S=-6$ sector, the binding of
$\Omega\Omega N$ and $\Omega\Omega NN$ depends strongly
on the coupling of the attractive
$\Omega N({}^{5}S_2)$ and $\Omega\Omega({}^{1}S_0)$
channels.
These results demonstrate that the formation of
multistrange bound states is governed not only by
the strength of the baryon--baryon interactions
but also by spin recoupling and the constraints
imposed by the Pauli principle.

The calculated spatial distributions reveal a compact
$\Omega$-centered structure in the strongly bound
$\Omega NN$ and $\Omega NNN$ states.
The $\Omega$ baryon remains closer to the center-of-mass
than the nucleons, with characteristic rms distances of
$0.8$--$0.9$~fm compared with approximately $1.2$~fm
for the nucleons.
Unlike the diffuse two-body $\Omega N$ system,
with an rms separation of about $4$~fm,
these few-body states form compact configurations.
The addition of a nucleon substantially increases
the binding without significantly changing the
spatial extent of the nucleon distribution.

Our results demonstrate that Coulomb interactions can significantly modify both the binding energies and spatial distributions of these exotic systems. Their effects depend on the charge composition and internal structure, with attractive $\Omega^-$–proton contributions competing against repulsive $\Omega^-$–$\Omega^-$ and proton–proton interactions. Although the predicted binding energies exhibit some dependence on the adopted $\Omega N$ potential, the qualitative features of the bound states remain largely unchanged. These findings provide new theoretical predictions for light $\Omega$-baryon nuclei and establish a microscopic few-body framework for further investigations of multistrange nuclear systems. More precise baryon–baryon interactions and future experimental constraints will be essential for assessing their possible existence and stability.

Since the pioneering studies of Witten \cite{Witten1984} and of Farhi and Jaffe 
in 1984~\cite{Jaffe1984}, the possible stability 
of strange quark matter has remained an important topic 
in theoretical physics~\cite{Zhang2024}. 
In contrast to bulk strange quark matter, whose stability 
is generally considered under conditions of high baryon 
density, light nuclei containing $\Omega$ baryons and 
nucleons may form bound states at much lower densities, 
potentially representing a distinct class of exotic 
multistrange nuclear matter.



\end{document}